\documentclass[]{iopart}

\expandafter\let\csname equation*\endcsname\relax
\expandafter\let\csname endequation*\endcsname\relax

\usepackage{graphicx} 
\usepackage{epstopdf}       
\usepackage{cite}
\usepackage[perpage,marginal]{footmisc}
\usepackage{etoolbox}

\usepackage{marvosym}
\usepackage{yfonts}
\usepackage{upgreek}
\usepackage{pdflscape}
\usepackage[unicode]{hyperref}
\usepackage[all]{hypcap}
\usepackage{amsmath,amsfonts,amssymb}
\usepackage{color}
\usepackage[usenames,dvipsnames]{xcolor}
\usepackage{bm}
\usepackage{longtable}
\usepackage[T1]{fontenc}
\usepackage[utf8]{inputenc}
\usepackage{verbatim}
\usepackage{pifont}
\usepackage{yfonts}
\usepackage{url}
\usepackage{multirow}
\usepackage{tensor}
\usepackage{mathtools}
\makeatletter
\def\url@leostyle{%
	\@ifundefined{selectfont}{\def\UrlFont{\sf}}{\def\UrlFont{\small\ttfamily}}}
\makeatother
\newcommand{\be}{\begin{equation}}
\newcommand{\ee}{\end{equation}}
\newcommand{\ba}{\begin{eqnarray}}
\newcommand{\ea}{\end{eqnarray}}

\makeatletter
\def\@mkboth#1#2{}
\newlength\appendixwidth
\preto\appendix{\addtocontents{toc}{\protect\patchl@section}}
\newcommand{\patchl@section}{%
	\settowidth{\appendixwidth}{\textbf{Appendix }}%
	\addtolength{\appendixwidth}{1.5em}%
	\patchcmd{\l@section}{1.5em}{\appendixwidth}{}{\ddt}%
}
\makeatother

\def\pd{\partial}
\def\pa{\partial}
\def\nn{\nonumber}

\usepackage[normalem]{ulem}
\usepackage{fancyhdr}

\makeatletter
\renewcommand\footnoterule{%
	\kern-3\p@
	\hrule\@width2.5cm
	\kern2.6\p@}
\makeatother

\numberwithin{equation}{section}
\numberwithin{figure}{section}

\newcommand{\sAE}{\ifmmode\text{\AE}\else\AE\fi}

\makeatletter
\renewcommand\@appendixstar{\@@par
	\ifnumbysec
	\@addtoreset{table}{section}
	\@addtoreset{figure}{section}\fi
	\setcounter{section}{0}
	\setcounter{subsection}{0}
	\setcounter{subsubsection}{0}
	\setcounter{equation}{0}
	\setcounter{figure}{0}
	\setcounter{table}{0}
	\def\thesection{Appendix \Alph{section}}
	\def\thesubsection{\Alph{subsection}}
	\def\theequation{\ifnumbysec
		\Alph{section}.\arabic{equation}\else
		\Alph{section}\arabic{equation}\fi}
	\def\thetable{\ifnumbysec
		\Alph{section}\arabic{table}\else
		A\arabic{table}\fi}
	\def\thefigure{\ifnumbysec
		\Alph{section}\arabic{figure}\else
		A\arabic{figure}\fi}}
\makeatother

\newcommand{\A}{{\scriptscriptstyle{A}}}
\newcommand{\B}{{\scriptscriptstyle{B}}}
\newcommand{\CC}{{\scriptscriptstyle{C}}}
\newcommand{\D}{{\scriptscriptstyle{D}}}
\newcommand{\E}{{\scriptscriptstyle{E}}}
\newcommand{\I}{{\scriptscriptstyle{I}}}
\newcommand{\J}{{\scriptscriptstyle{J}}}
\newcommand{\M}{{\scriptscriptstyle{M}}}
\newcommand{\N}{{\scriptscriptstyle{N}}}
\newcommand{\K}{{\scriptscriptstyle{K}}}
\newcommand{\LL}{{\scriptscriptstyle{L}}}

\begin{document}
\newcounter{count}

	\pagestyle{fancy}
	\lhead{GW extraction in higher dimensional NR using the Weyl tensor}
	\chead{}
	\rhead{\thepage}
	\lfoot{}
	\cfoot{}
	\rfoot{}
	
	\begin{center}
	\title{Gravitational wave extraction in higher dimensional numerical relativity using the Weyl tensor}
	\end{center}
	
\author{%
	William G. Cook$^{1}$,
	Ulrich Sperhake$^{1,2,3}$
}
\address{$^{1}$~Department of Applied Mathematics and Theoretical Physics, Centre for Mathematical Sciences, University of Cambridge,
	Wilberforce Road, Cambridge CB3 0WA, United Kingdom}
\address{$^{2}$~Department of Physics and Astronomy,
	The University of Mississippi, University, MS 38677-1848, USA}
\address{$^{3}$~Theoretical Astrophysics 350-17,
	California Institute of Technology, Pasadena, CA 91125, USA}

\ead{wc259@damtp.cam.ac.uk}

\begin{abstract}
Gravitational waves are one of the most important diagnostic
tools in the analysis of strong-gravity dynamics and have been
turned into an observational channel with
LIGO's detection of GW150914. Aside from their
importance in astrophysics, black holes and compact matter
distributions have also assumed a central role in many other
branches of physics. These applications often involve
spacetimes with $D>4$ dimensions where the calculation of gravitational
waves is more involved than in the
four dimensional case, but has now become possible thanks to
substantial progress in the theoretical study of
general relativity in $D>4$.
Here, we develop a numerical implementation of the formalism
by Godazgar \& Reall \cite{Godazgar:2012zq} -- based on projections
of the Weyl tensor analogous to the Newman-Penrose scalars -- that allows for
the calculation of gravitational waves in higher dimensional
spacetimes with rotational symmetry. We apply and test this method in
black-hole head-on collisions from rest in $D=6$ spacetime dimensions
and find that a fraction $(8.19\pm 0.05)\times 10^{-4}$
of the Arnowitt-Deser-Misner mass is radiated away from the system,
in excellent agreement with literature results based on
the Kodama-Ishibashi perturbation technique. The method
presented here complements the perturbative approach by
automatically including contributions from all multipoles
rather than computing the energy content of individual multipoles.

\end{abstract}

\section{Introduction}

Gravitational waves (GWs) entered the limelight with the recent
detection of GW150914 \cite{Abbott:2016blz} -- soon followed
by a second detection GW151226 \cite{Abbott:2016nmj} --
which not only constitutes
the first observation of a black-hole (BH) binary system, but also
marks a true milestone in gravitational physics. This breakthrough has opened a
quantitatively new path for
measuring BH parameters \cite{TheLIGOScientific:2016wfe,Abbott:2016apu},
testing Einstein's theory of general relativity
\cite{TheLIGOScientific:2016src} and probing extreme astrophysical
objects and their formation history
\cite{TheLIGOScientific:2016htt}, and substantially
broadens the scope of multi-messenger astronomy
\cite{Adrian-Martinez:2016xgn}. GW modelling,
however, finds important applications beyond the
revolutionary field of GW astronomy. Many fundamental questions
in general relativity in $D=4$ and $D>4$ spacetime
dimensions concern the stability of strong-gravity sources (see
\cite{Regge:1957td,Zerilli:1970se,Emparan:2008eg,Figueras:2015hkb,Shibata:2010wz,Santos:2015iua,
Zilhao:2014wqa,Dafermos:2010hd} and references therein)
in the context of cosmic censorship violation, the solutions'
significance as physical objects or expanding our understanding
of the strong-field regime of general relativity.
For instance, GW extraction from numerical simulations of
rapidly spinning Myers-Perry BHs demonstrates how excess angular
momentum is shed in order to allow the BH to settle down into
a more slowly rotating configuration \cite{Shibata:2010wz}.
GW emission also represents a channel for mass-energy loss
in ultra-relativistic collisions that are studied in the
context of the so-called TeV gravity scenarios that may
explain the hierarchy problem in physics;
for reviews see e.g.~\cite{Kanti:2008eq,Cardoso:2014uka,Choptuik:2015mma}.

The calculation of GW signals in the theoretical modelling of
$D=4$ dimensional sources in the framework of general relativity
has been increasingly well understood following seminal work
by Pirani, Bondi, Sachs and others
in the 1950s and 1960s; see e.g.~\cite{Bondi:1958aj,Pirani:1956wr,Bondi:1957dt,Bondi:1960jsa,Newman:1961qr,Bondi:1962px,Sachs:1962wk} and \cite{Saulson:2010zz}
for a review. Applications are now routinely found in numerical
and (semi-)analytic calculations \cite{Peters:1964zz,Cunningham:1978zfa,Cunningham:1979px,Blanchet:1989fg,Blanchet:1995ez,Ruiz:2007yx,Nerozzi:2008ng,Centrella:2010mx,Reisswig:2010cd,Hinder:2013oqa}
even though care needs to be taken
when applied to numerical simulations on finite domains
\cite{Lehner:2007ip}.

The numerical study of solutions to Einstein's equations
has proven incredibly useful for understanding the
behaviour of black holes and other compact objects.
Most recently, the application of numerical relativity in the
generation of gravitational waveform templates for GW data analysis
\cite{Read:2013zra,Hinder:2013oqa,
Mroue:2013xna,Taracchini:2013rva,Purrer:2014fza,Khan:2015jqa,Hannam:2013oca}
contributed to the above mentioned detection of GW150914.

The wave extraction techniques presently used in numerical simulations
of astrophysical GW sources can be classified as follows:
perturbative methods based on the formalism
developed by Regge, Wheeler, Zerilli and Moncrief \cite{Regge:1957td,
Zerilli:1970se, Moncrief:1974am};
application of the quadrupole formula \cite{Thorne1980} in
matter simulations \cite{Ott:2006eu,Shibata:2004kb};
a method using the Landau-Lifshitz
pseudo-tensor \cite{1975ctfbookL, Lovelace:2009dg} ; Cauchy
characteristic extraction \cite{Reisswig:2009us,Reisswig2009a,Babiuc:2010ze};
and, probably the most popular technique, using
the Weyl scalars from the Newman-Penrose formalism \cite{Newman:1961qr,
Pretorius:2005gq,Baker:2005vv,Campanelli:2005dd,Sperhake:2006cy,Herrmann:2007ac,
Brugmann:2008zz,Boyle:2007ft}. For a comparison of various of these
techniques, see \cite{Reisswig:2010cd}.

The calculation of GWs in higher dimensional relativity requires
generalization of these techniques to $D>4$. The extraction of the GW
energy flux from the Landau-Lifshitz pseudotensor has been generalized
straightforwardly to higher dimensions in \cite{Cardoso:2002pa,Yoshino:2009xp}.
An extension of the Regge-Wheeler-Zerilli-Moncrief formalism for
perturbations of spherically symmetric background spacetimes
is available in the form of the Kodama and Ishibashi (KI) formalism
\cite{Kodama:2003jz,Ishibashi:2011ws} and forms the basis
of the wave extraction techniques developed in
\cite{Witek:2010xi,Witek:2011zz}. The main assumption there is that
far away from the strong-field regime, the spacetime is
perturbatively close to the Tangherlini \cite{Tangherlini:1963bw}
spacetime. The deviations from this background facilitate
the construction of master functions according to the KI formalism
which in turn provide the energy flux in the different
$(l,m)$ radiation multipoles. Even though both methods are
in practice applied at finite extraction radius, their
predictions have been found to agree within a $\sim1~\%$ error
tolerance when applied to BH head-on collisions starting from rest
in $D=5$ \cite{Witek:2014mha}. Recent years have also seen considerable
progress in the understanding of the peeling properties of the
Weyl tensor; see \cite{Godazgar:2012zq,Ortaggio:2012hc} and references
therein.

In particular, Godazgar \& Reall
\cite{Godazgar:2012zq} have performed a
decomposition of the Weyl tensor in
higher dimensions, and derived a generalisation of the
Newman-Penrose formalism for wave extraction to $D>4$.
The numerical implementation of this formalism
and probing the accuracy for a concrete example application
is the subject of this paper.

For this purpose, we require a formulation of the higher dimensional
Einstein equations suitable for numerical integration.
For computational practicality, all higher dimensional time
evolutions in numerical relativity have employed symmetry assumptions
to reduce the problem to an effective ``$d$+1'' dimensional
computation where typically $d\le 3$.
This has been achieved in practice by either
(i) imposing the symmetries directly on the metric
line element \cite{Chesler:2013lia}, (ii) dimensional
reduction by isometry of the Einstein field equations
\cite{Zilhao:2010zz,Zilhao:2013gu} or (iii) use of
so-called {\em Cartoon} methods
\cite{Pretorius:2004jg,Shibata:2009ad,Shibata:2010wz,Yoshino:2011zz,Yoshino:2011zza}.
In our implementation, we use the latter method combined with
the Baumgarte-Shapiro-Shibata-Nakamura
\cite{Shibata:1995we,Baumgarte:1998te}
(BSSN) formulation of
the Einstein equations as expanded in detail in
\cite{Cook:2016soy}. The relevant expressions for the
GW computation, however, will be expressed in terms of the
Arnowitt-Deser-Misner \cite{Arnowitt:1962hi} (ADM) variables,
and the formalism as presented here can be straightforwardly
applied in other common evolution systems
used in numerical relativity.

The paper is structured as follows. In Section \ref{sec:notation}
we describe the notation used in the remainder of this work. In
Section \ref{sec:Theory} we recapitulate
the key results of \cite{Godazgar:2012zq}
which sets up the formalism. In Section \ref{sec:MC} we put the
formalism into a form compatible with the modified Cartoon dimensional
reduction of our simulations. In Section \ref{sec:Numerical} we
describe the numerical set up used in our simulations,
analyse the energy radiated in BH collisions in $D=6$
and compare the predictions with literature results
based on alternative wave extraction techniques.

\section{Notation and Indices} \label{sec:notation}
The key goal of our work is to extract the GW signal from dynamical,
asymptotically flat
$D$ dimensional spacetimes, i.e.~manifolds $\mathcal{M}$
equipped with a metric $g_{\A \B}$, $A,\,B = 0,\,\ldots,\,D-1$,
of signature $D-2$
that obeys the Einstein equations with vanishing cosmological constant
\begin{equation}
  G_{\A\B}=R_{\A\B}-\frac{1}{2}Rg_{\A \B} = 8\pi T_{\A \B}\,.
  \label{eq:EinsteinD}
\end{equation}
Here, $T_{\A \B}$ is the energy momentum tensor which we assume to
vanish in the wave zone, $R_{\A\B}$ and $R$ denote the Ricci tensor and
scalar, respectively, and we are using units where the
gravitational constant and the speed of light $G=c=1$. Furthermore,
we define the Riemann tensor and Christoffel symbols according to
the conventions of Misner, Thorne \& Wheeler \cite{Misner1973}.

The ADM space-time decomposition as reformulated by York
\cite{York1979} writes the metric line element in the form
\begin{equation}
  ds^2 = g_{\A\B}dx^{\A}dx^{\B}
        = (-\alpha^2 + \beta_{\I}\beta^{\I})dt^2
        + 2\beta_{\I} dx^{\I} dt + \gamma_{\I\J} dx^{\I}dx^{\J}\,,
  \label{eq:ADMlineelement}
\end{equation}
where $I,\,J = 1,\,\ldots,\,D-1$ and $\alpha$, $\beta^{\I}$ denote the
lapse function and shift vector, respectively, and $\gamma_{\I\J}$
is the spatial metric that determines the geometry of hypersurfaces
$t={\rm const}$. For this choice of coordinates and variables,
the Einstein equations (\ref{eq:EinsteinD}) result in one
{\em Hamiltonian} and $D-1$ {\em momentum} constraints as well
as $D(D-1)/2$ evolution equations cast into first-order-in-time form
by introducing the extrinsic curvature $K_{\I\J}$ through
\begin{equation}
  \partial_t \gamma_{\I\J} = \beta^{\M}\partial_{\M} \gamma_{\I\J}
        + \gamma_{\M\J} \partial_{\I} \beta^{\M}
        + \gamma_{\I\M} \partial_{\J} \beta^{\M}
        - 2\alpha K_{\I\J}\,;
        \label{eq:EK}
\end{equation}
for a detailed review of this decomposition see \cite{Gourgoulhon:2007ue,
Sperhake:2013wva}.

In the remainder of this work we assume that the ADM variables are
available. The BSSN formulation, for example, employs
a conformally rescaled spatial metric $\tilde{\gamma}_{\I\J}$,
the rescaled traceless extrinsic curvature $\tilde{A}_{\I\J}$,
a conformal factor $\chi$, the trace of the extrinsic curvature $K$
and contracted Christoffel symbols $\tilde{\Gamma}^{\I}$ associated
with the conformal metric. These are related to the ADM variables
by
\begin{align}
  \chi = \gamma^{-1/(D-1)}\,,
  ~~~~~
  & K = \gamma^{\M \N} K_{\M \N}, \nonumber \\
  \tilde{\gamma}_{\I \J} = \chi \gamma_{\I \J}
  ~~~
  &\Leftrightarrow
  ~~~
  \tilde{\gamma}^{\I \J} = \frac{1}{\chi} \gamma^{\I \J}, \nonumber \\
  \tilde{A}_{\I \J} = \chi \left( K_{\I \J} - {\displaystyle \frac{1}{D-1}}
  \gamma_{\I \J} K \right)
  ~~~
  &\Leftrightarrow
  ~~~
  K_{\I \J} = \frac{1}{\chi} \left( \tilde{A}_{\I \J} + \frac{1}{D-1}
  \tilde{\gamma}_{\I \J} K \right)\,, \nonumber \\[10pt]
  \tilde{\Gamma}^{\I} = \tilde{\gamma}^{\M \N}
  \tilde{\Gamma}^{\I}_{\M \N}\,,&
  \label{eq:BSSNvars}
\end{align}
so that the ADM variables can be reconstructed straightforwardly at
every time in the evolution.
For other popular evolution systems such as
the conformal Z4 system \cite{Alic:2011gg,Hilditch:2012fp} or
the generalized harmonic formulation
\cite{Garfinkle:2001ni,Pretorius:2004jg}
the ADM variables are obtained in similar fashion or directly from the
spacetime metric components through
Eqs.~(\ref{eq:ADMlineelement}), (\ref{eq:EK}).

Many practical applications of higher dimensional numerical relativity
employ symmetry assumptions that reduce the computational domain to
an effective three dimensional spatial grid using the
aforementioned techniques. The reasons are two-fold: (i) the
computational cost of simulations in $D>3+1$ dimensions massively
increase with any dimension added and (ii) the $SO(D-d)$ rotational
symmetry
accomodates many
applications of interest that can therefore be handled by relatively
straightforward extensions of numerical codes originally developed
for astrophysical systems in four spacetime dimensions.

In general, the number of effective (spatial) dimensions can
range inside $1\le d \le D-2$, where $d=1$ corresponds to spherical
symmetry, i.e.~$SO(D-1)$ isometry, and the other extreme $d=D-2$
corresponds to the axisymmetric case $SO(2)$ with rotational
symmetry about one axis. As already detailed in \cite{Cook:2016soy},
axisymmetry, i.e.~$d=D-2$, represents a special case that
imposes fewer constraints on the components of tensors and their
derivatives and is therefore most conveniently handled numerically
in a manner similar but not identical to the general case.
We will discuss first in detail the case $1\le d\le D-3$
and then describe the specific recipe for dealing with $d=D-2$.
Probably the most important situation encountered in
practical applications is that of an effective $d=3$ dimensional
spatial computational domain. Whenever the expressions
developed in the remainder of this work for general $d$
are not
trivially obtained for the special choice $d=3$,
we shall therefore explicitly write down the $d=3$ version
in addition to the general case.

Let us start by considering a spacetime with at least
two rotational Killing vectors which is the scenario
discussed
in Secs.~2 and 3 of Ref.~\cite{Cook:2016soy}.
For convenience, we will employ two
specific coordinate systems adapted to
this situation. The first is a set of Cartesian coordinates
\begin{equation}
  X^{\A} = (t,\,\underbrace{x^1,\,\ldots,x^{d-1}}_{(d-1)\times},\,
        z,\,\underbrace{w^{d+1},\,\ldots,\,w^{D-1}}_{(D-d-1)\times}
        = (t,\,x^{\hat{i}},\,z,\,w^a)=(t,\,x^i,\,w^a)\,,
  \label{eq:coords_Cart}
\end{equation}
where middle Latin indices without (with) a caret range from
$i=1,\,\ldots,\,d$ ($\hat{i} = 1,\,\ldots,\,d-1$),
early Latin indices run from
$a=d+1,\ldots,D-1$, and rotational Killing vector fields are
presumed to exist in all planes spanned by any two of the coordinates
$(z,w^a)$. The second is a system of spherical coordinates
\begin{equation}
  Y^{\A} = (t,\,r,\,\underbrace{\phi^2,\,\phi^3,\,\ldots,\,
        \phi^{D-1}}_{(D-2)\times}\,) = (t,\,r,\,\phi^{\alpha})\,,
  \label{eq:coords_spher}
\end{equation}
where Greek indices run from $\alpha=2,\ldots,D-1$.
We use middle, upper case Latin indices to denote all spatial coordinates
of either of these systems, so that $X^{\I} = (x^{\hat{i}},\,z,\,w^a)$
and $Y^{\I} = (r,\,\phi^{\alpha})$ with $I = 1,\ldots,D-1$.
In the special case $d=3$, we use the notation
$x^{\hat{i}} \equiv (x,\,y)$, so that Eq.~(\ref{eq:coords_Cart}) becomes
\begin{equation}
  X^{\A} = (t,\,x,\,y,\,z,\,w^4,\,\ldots,\,w^{D-1})\,.
  \label{eq:coords_Cartd3}
\end{equation}

We orient the Cartesian coordinates (\ref{eq:coords_Cart})
such that they are related to the spherical coordinates
(\ref{eq:coords_spher}) by
\begin{align}
  (w^{1}\equiv)~~x^{1} &= r\cos \phi^2 \,, \nonumber \\[10pt]
  (w^{2}\equiv)~~x^{2} &= r\sin \phi^2\,\cos \phi^3\,, \nonumber \\
  \vdots & \nonumber \\[0pt]
  (w^{d-1}\equiv)~~x^{d-1} &= r\sin \phi^2\,\ldots\,\sin
        \phi^{d-1}\,\cos\phi^d\,, \nonumber \\[10pt]
  (w^d\equiv)~~z &= r\sin \phi^2\,\ldots\,\sin \phi^{d-1}\,
        \sin \phi^d\,\cos \phi^{d+1}\,,
        \nonumber \\[10pt]
  w^{d+1} &= r\sin \phi^2\,\ldots\,\sin \phi^{d-1}\,
        \sin \phi^d\,\sin \phi^{d+1}\,\cos \phi^{d+2}\,,
        \nonumber \\
  \vdots \nonumber \\[10pt]
  w^{D-3} &= r\sin \phi^2\,\ldots\,\sin \phi^{D-3}\,\cos \phi^{D-2} \,,
        \nonumber \\[10pt]
  w^{D-2} &= r\sin \phi^2\,\ldots\,\sin \phi^{D-3}\,\sin \phi^{D-2}
        \,\cos \phi^{D-1}\,, \nonumber \\[10pt]
  w^{D-1} &= r\sin \phi^2\,\ldots\,\sin \phi^{D-3}\,\sin \phi^{D-2}
        \,\sin \phi^{D-1}\,.
  \label{eq:coordinates}
\end{align}
Here $\phi^{D-1}\in [0,2\pi]$, and all other $\phi^{\alpha}\in[0,\pi]$,
and we have formally extended the $w$ coordinates to also include
(in parentheses in the equation)
$w^i = x^i$ which turns out to be convenient in the notation below in
Sec.~\ref{sec:MC}.
Note that for the orientation chosen here,
the $x$ axis denotes the reference axis for the
first polar angle rather than the $z$ axis which more commonly plays this role
for spherical coordinates in three spatial dimensions.

For orientation among the different sets of indices, we conclude this section
with a glossary listing index variables and their ranges as employed
throughout this work.
\begin{itemize}
  \item Upper case early Latin indices $A,\,B,\,C,\,\ldots$ range over the
        full spacetime from 0 to $D-1$.
  \item Upper case middle Latin indices $I,\,J,\,K,\,\ldots$ denote all
        spatial indices, inside and outside the effective three dimensional
        computational domain, running from 1 to $D-1$.
  \item Lower case middle Latin indices $i,\,j,\,k,\,\ldots$ denote indices
        in the spatial computational domain and run from $1$ to $d$.
        For $d=3$, we have $x^i=(x,\,y,\,z)$.
  \item Lower case middle Latin indices with a caret
        $\hat{i},\,\hat{j},\,\ldots$ run from $1$ to $d-1$ and
        represent the $x^i$ (without caret) excluding $z$. In
        $d=3$, we write $x^{\hat{i}}=(x,\,y)$.
  \item Lower case early Latin indices $a,\,b,\,c,\,\ldots$ denote
        spatial indices outside the computational domain, ranging from $d+1$ to $D-1$.
  \item Greek indices $\alpha,\,\beta,\,\ldots$
        denote all angular directions and range from $2$ to $D-1$.
  \item Greek indices with a caret $\hat{\alpha}, \hat{\beta},\,\ldots$
        denote the subset of angular coordinates in the computational
        domain, i.e.~range from $2,\,\ldots,\,d$.
        As before, a caret thus indicates a truncation of the index range.
  \item Put inside parentheses, indices cover the same range but
        merely denote labels rather than tensor indices.
  \item An index 0 denotes a contraction with the timelike normal
        to the foliation, rather than the time component, as detailed
        below in Section \ref{D-1+1split}.
  \item $\nabla_{\A}$ denotes the covariant derivative in the full
        $D$ dimensional spacetime, whereas $D_{\I}$ denotes the covariant
        derivative on a spatial hypersurface and is calculated from
        the spatial metric $\gamma_{\I\J}$.
  \item We denote by $R$ with appropriate indices the Riemann tensor
        (or Ricci tensor/scalar) of the full $D$ dimensional spacetime, and
        by $\mathcal{R}$ the spatial Riemann tensor (or Ricci
        tensor/scalar) calculated from $\gamma_{\I \J}$.
\end{itemize}

\section{Theoretical Formalism} \label{sec:Theory}
Our wave extraction from numerical BH simulations in $D>4$ dimensions
is based on the formalism developed by Godazgar \& Reall
\cite{Godazgar:2012zq}. In this section, we summarize the key findings and
expressions from their work.

The derivation \cite{Godazgar:2012zq}
is based on the definition of asymptotic flatness
using Bondi coordinates \cite{Sachs:1962wk}
$(u,\textswab{r},\phi^{\alpha})$
where $u$ is retarded time, $\textswab{r}$ the radius
and $\phi^{\alpha}$ are
$D-2$ angular coordinates covering the unit $D-2$ sphere.
A spacetime is asymptotically flat
at future null infinity \cite{Tanabe:2011es}
if the metric outside a cylindrical world tube of
finite radius can be written in terms of functions
$\mathcal{A}(u,\textswab{r},\phi^{\alpha})$,
$\mathcal{B}(u,\textswab{r},\phi^{\alpha})$,
$\mathcal{C}(u,\textswab{r},\phi^{\alpha})$
as
\begin{equation}
  ds^2 = -\mathcal{A}e^{\mathcal{B}} du^2 -2e^{\mathcal{B}}du\,
        d\textswab{r} + \textswab{r}^2
        h_{\alpha\beta}(d\phi^{\alpha}
        +\mathcal{C}^{\alpha}du)(d\phi^{\beta}
        +\mathcal{C}^{\beta}du)\,,
\end{equation}
with $\det h_{\alpha\beta}=\det\omega_{\alpha\beta}$
where $\omega_{\alpha\beta}$ is the unit metric on the $D-2$ sphere.
For an asymptotically flat spacetime $h_{\alpha\beta}$
can be expanded as \cite{Tanabe:2011es}
\begin{equation}
  h_{\alpha\beta}=\omega_{\alpha\beta}(\phi^{\gamma})+\sum_{s\geq0}
        \frac{h_{\alpha\beta}^{(s+1)}(u,
        \phi^{\gamma})}{\textswab{r}^{D/2+s-1}}\,,
\end{equation}
and the Bondi news function is obtained from this expansion
as the leading-order correction $h^{(1)}_{\alpha\beta}$.

In analogy with the $D=4$ case, a null frame of vectors is constructed
which is asymptotically given by
\footnote{The construction of the exact analog of the
Kinnersley \cite{Kinnersley:1969zza} tetrad in general spacetimes
at finite radius is subject of ongoing research even in $D=4$
(see e.g.~\cite{Nerozzi:2008ng,Zhang:2012ky}). In practice,
the error arising from the use of an asymptotic form
of the tetrad at finite extraction radii is mitigated
by extracting at various radii and extrapolating to
infinity \cite{Hinder:2013oqa} and we pursue this approach,
too, in this work.}

\begin{align}
  l=-\frac{\partial}{\partial \textswab{r}}, ~\,~
  k= \frac{\partial}{\partial u} - \frac{1}{2}
        \frac{\partial}{\pa \textswab{r}}, ~\,~
  m_{(\alpha)} =\frac{\pa}{\pa \phi^{\alpha}}\,.
  \label{eq:tetrad}
\end{align}
Note that all the tetrad vectors are real in contrast to the $D=4$
dimensional case where the two vectors $m_{(2)}$ and $m_{(3)}$ are
often written as two complex null vectors.
Next, the components of the Weyl tensor are projected onto the frame
(\ref{eq:tetrad})
and the leading order term
in the radial coordinate is extracted.
Following \cite{Godazgar:2012zq}, we denote this quantity by
$\Omega'$ and its components are given by
\begin{equation}
  \Omega'_{(\alpha)(\beta)}\equiv C_{\A\B\CC\D}
        k^{\A} m_{(\alpha)}^{\B}k^{\CC} m_{(\beta)}^{\D}
        = - \frac{1}{2} \frac{\hat{e}_{(\alpha)}^{\mu}
        \hat{e}_{(\beta)}^{\nu}\ddot{h}^{(1)}_{\mu\nu}}
        {\textswab{r}^{D/2-1}}+\mathcal{O}(\textswab{r}^{-D/2})\,.
  \label{eq:weylcomps}
\end{equation}
Here $\hat{e}_{(\alpha)}^{\beta}$ denote a set of vectors
forming an orthonormal basis for the unit metric
$\omega_{\alpha\beta}$ on the $D-2$ sphere. In practice, this basis
is constructed using Gram-Schmidt orthonormalisation starting with
the radial unit vector.

As with the Newman-Penrose scalar
$\Psi_4$ in the four dimensional case,
we note that this is the contraction of the Weyl
tensor with the ingoing null vector twice and two spatial vectors.
Whereas in $D=4$ the two polarisations of the gravitational waves are
encoded in the real and imaginary parts of $\Psi_4$,
here $\Omega_{(\alpha)(\beta)}'$
is purely real, with the $\alpha,\,\beta$ labels providing the different
polarisations.

The final ingredient for extracting the energy radiated in GWs
is the rate of change of the Bondi mass given by \cite{Tanabe:2011es}
\begin{equation}
  \dot{M}(u)=\frac{1}{32\pi}\int_{S^{D-2}}\dot{h}^{(1)}_{\alpha\beta}
        \dot{h}^{(1)\alpha\beta} d\omega\,.
\end{equation}
By substituting in for $\dot{h}^{(1)}_{\alpha\beta}$
from the definition of $\Omega'_{(\alpha)(\beta)}$ we obtain an expression
for the mass loss.
\begin{equation}
  \dot{M}(u)=-\lim_{\textswab{r}\rightarrow \infty}
        \frac{\textswab{r}^{D-2}}{8\pi}\int_{S^{D-2}}\left(\int^u_{-\infty}
        \Omega'_{(\alpha)(\beta)}
        (\tilde{u},\textswab{r},\phi^{\gamma})d\tilde{u}\right)^2
        d\omega\,, \label{Mdot}
\end{equation}
where the notation $(\ldots)^2$
implies summation over the $(\alpha),\,(\beta)$ labels inside the parentheses,
and $d\omega$ denotes the area element of the $D-2$ sphere.
In practice, we will apply Eq.~(\ref{Mdot}) at constant radius $\textswab{r}$,
therefore replace retarded time $u$ with ``normal'' time $t$ and start
the integration at $t=0$ rather than $-\infty$,
assuming that GWs generated prior to the start of the simulation can be
neglected.

\section{Modified Cartoon Implementation} \label{sec:MC}

The formalism summarized in the previous section is valid in
generic $D$ dimensional spacetimes with or without symmetries.
We now assume that the spacetime under consideration obeys
$SO(D-d)$ isometry with $1\le d\le D-3$,
and will derive the expressions required for
applying the GW extraction formalism of Sec.~\ref{sec:Theory}
to numerical simulations
employing the modified Cartoon method.

Throughout this derivation, we will make use of the expressions
for scalars, vectors and tensors in spacetimes with
$SO(D-d)$ symmetry and the regularisation of their components
at $z=0$ as listed in Appendices A and B of Ref.~\cite{Cook:2016soy}.
The key result of these relations for our purposes is that the ADM
variables $\alpha$, $\beta^{\I}$, $\gamma_{\I\J}$, $K_{\I\J}$
for a spacetime with $SO(D-d)$ isometry can be expressed completely
in terms of their $d$ dimensional components $\beta^i$, $\gamma_{ij}$
and $K_{ij}$ as well as two additional functions $\gamma_{ww}$
and $K_{ww}$ according to
\begin{align}
  \beta^{\I} &= (\beta^i, 0)\,, \nonumber \\
  \gamma_{\I\J} &= \left(
        \begin{matrix}
          \gamma_{ij} & 0 \\
          0 & \delta_{ab} \gamma_{ww}
        \end{matrix} \right)\,, \nonumber \\
  K_{\I\J} &= \left(
        \begin{matrix}
          K_{ij} & 0 \\
          0 & \delta_{ab} K_{ww}
        \end{matrix}\right)\,,
  \label{eq:ADMSO3}
\end{align}
while the scalar $\alpha$ remains unchanged.

From the viewpoint of numerical applications, the key relations of
the procedure reviewed in Sec.~\ref{sec:Theory} are
Eqs.~(\ref{eq:weylcomps}) and (\ref{Mdot}). The first provides
$\Omega_{(\alpha)(\beta)}'$ in terms of the Weyl tensor and the normal frame,
and the second tells us how to calculate the mass loss from
$\Omega_{(\alpha)(\beta)}'$. The latter is a straightforward integration
conveniently applied as a
post processing operation, so that we can focus here on the former
equation. For this purpose, we first note that in practice
wave extraction is performed in the wavezone far away from the sources.
Even if the sources are made up of non-trivial energy matter fields,
the GW signal is calculated in vacuum where the Weyl and Riemann tensors
are the same. Our task at hand is then twofold: (i) calculate the Riemann
tensor from the ADM variables and (ii) to construct a null frame.
These two tasks are the subject of the remainder of this section.

\subsection{The Riemann Tensor}

\subsubsection{The $(D-1)+1$ splitting of the Riemann tensor} \label{D-1+1split} 

\paragraph{}
The ADM formalism is based on a space-time decomposition of the $D$
dimensional spacetime manifold into a one-parameter family of spacelike
hypersurfaces which are characterized by a future-pointing, unit normal
timelike field $n^{\A}$. This normal field together with the projection
operator
\begin{equation}
  \bot^{\A}{}_{\B}=\delta^{\A}{}_{\B}+n^{\A} n_{\B}\,,
\end{equation}
allows us to split tensor fields into components tangential or orthogonal
to the spatial hypersurfaces by contracting each tensor index either with
$n_{\A}$ or with $\bot^{\B}{}_{\A}$. For a symmetric rank (0,2) tensor,
for example, we thus obtain the following three contributions
\begin{align}
  &T_{00} \equiv T_{\A\B}n^{\A} n^{\B}\,,~~~~~
  \bot T_{0\A} = \bot T_{\A 0}\equiv \bot^{\CC}{}_{\A} T_{\CC \B}n^{\B}\,,~~~~~~
  \bot T_{\A\B} \equiv \bot^{\CC}{}_{\A} \bot^{\D}{}_{\B}T_{\CC\D}\,.
\end{align}

The most important projections for our study are those of the Riemann tensor
which are given by the Gauss-Codazzi relations used in the standard
ADM splitting of the Einstein Equations (see e.g.~\cite{Gourgoulhon:2007ue})
\begin{align}
  \bot R_{\A\B\CC\D} =&
        \mathcal{R}_{\A\B\CC\D}+K_{\A\CC}K_{\B\D}-K_{\A\D}K_{\CC\B},
  \label{Gauss} \\
  \bot R_{\A 0\CC\D} \equiv \bot (R_{\A\B\CC\D}n^{\B})=&
        -D_{\CC}K_{\A\D}+D_{\D}K_{\A\CC}, \label{Codazzi} \\
  \bot R_{\A 0\CC 0} \equiv \bot (R_{\A\B\CC\D}
        n^{\B} n^{\D})  =&
        \bot R_{\A\CC} + \mathcal{R}_{\A\CC}+KK_{\A\CC}-K_{\A\E}K^{\E}{}_{\CC}
        \nonumber \\
        =& \mathcal{R}_{\A\CC}+KK_{\A\CC}-K_{\A\E}K^{\E}{}_{\CC},
  \label{contrGauss}
\end{align}
where in the last line we used the fact that in vacuum $R_{\A\CC}$ and,
hence, its projection vanishes (note, however, that in general
$\mathcal{R}_{\A\CC}\ne 0$ even in vacuum).
Furthermore $D_{\CC}K_{\A\D}=\partial_{\CC}K_{\A\D}-\Gamma^{\B}_{\CC\A}K_{\B\D}
-\Gamma^{\B}_{\CC\D}K_{\A\B}$
is the covariant derivative of the extrinsic curvature defined on
the spatial hypersurface, with Christoffel symbols calculated from
the induced metric $\gamma_{\A\B}$.
Equations (\ref{Gauss})-(\ref{contrGauss}) tell us how to
reconstruct the full $D$ dimensional Riemann tensor from $D-1$ dimensional
quantities defined on the spatial hypersurfaces which
foliate our spacetime.

From this point on, we will use coordinates adapted to the $(D-1)+1$ split.
In such coordinates, we can replace in Eqs.~(\ref{Gauss})-(\ref{contrGauss})
the spacetime indices $A,\,B,\,\ldots$ on the left and right-hand side
by spatial indices
$I,\,J,\,\ldots$ while the time components of the spacetime
Riemann tensor are taken into account through the contractions
with the unit timelike normal $n^{\A}$ and which we denote with
the suffix $0$ in (\ref{Codazzi}),\,(\ref{contrGauss}). Note that
more than two contractions of the Riemann tensor with the
timelike unit normal $n^{\A}$ vanish by symmetry of the Riemann tensor.

\subsubsection{The Riemann tensor in $SO(D-3)$ symmetry}
\label{sec:Riemann}
\paragraph{}
The expressions given in the previous subsection for the components
of the Riemann tensor are valid for general spacetimes with or without
symmetries.
In this section, we
will work out the form of the components of the Riemann tensor in
spacetimes with $SO(D-d)$ isometry for $1\le d\le D-3$.

For this purpose we recall
the Cartesian coordinate system
$X^{\I}=(x^{\hat{i}},\,\,z,\,w^a)$
of Eq.~(\ref{eq:coords_Cart}),
adapted to a spacetime that is
symmetric under rotations in any plane spanned
by two of the $(z,\,w^a)$. We discuss in turn
how the terms appearing on the right-hand
sides of Eqs.~(\ref{Gauss})-(\ref{contrGauss}) simplify under this symmetry.
We begin with the components of the
spatial Riemann tensor, given in terms of the spatial metric
and Christoffel symbols by
\begin{align}
  \mathcal{R}_{\I\J\K\LL}=&\frac{1}{2}\left(\partial_{\LL}\partial_{\I}
        \gamma_{\J\K} +\partial_{\K}\partial_{\J}\gamma_{\I\LL}
        -\partial_{\K}\partial_{\I}\gamma_{\J\LL}
        -\partial_{\LL}\partial_{\J}\gamma_{\I\K}\right)\nn\\
  &-\gamma_{\M\N}\Gamma^{\N}_{\I\K}\Gamma^{\M}_{\J\LL}
        +\gamma_{\M\N}\Gamma^{\N}_{\I\LL}\Gamma^{\M}_{\J\K}\,.
  \label{eq:spatialRiemann}
\end{align}
The rotational symmetry imposes conditions on the derivatives of
the metric, the Christoffel symbols and the components of the
Riemann tensor that are obtained in complete analogy to the derivation
in Sec.~2.2 and Appendix A of \cite{Cook:2016soy}.
We thus calculate all components of the Riemann tensor,
where its indices can vary over the coordinates inside and outside
the computational domain, and obtain
\begin{align}
  \mathcal{R}_{ijkl} = & \frac{1}{2}\left(\partial_l\partial_i\gamma_{jk}
        +\partial_k\partial_j\gamma_{il}
        -\partial_k\partial_i\gamma_{jl}
        -\partial_l\partial_j\gamma_{ik}\right)
        -\gamma_{mn}\Gamma^n_{ik}\Gamma^m_{jl}
        +\gamma_{mn}\Gamma^n_{il}\Gamma^m_{jk},
        \label{eq:R3ijkl} \\
  \mathcal{R}_{ajkl} = & 0, \\
  \mathcal{R}_{iajb} = & \delta_{ab}\mathcal{R}_{iwjw}, \\
  \mathcal{R}_{iwjw} \equiv & \frac{\partial_{(i}\gamma_{j)z}
        -\delta_{z(j}\partial_{i)}\gamma_{ww}}{z}
        -\delta_{z(i}\frac{\gamma_{j)z}
        -\delta_{j)z}\gamma_{ww}}{z^2}
        -\frac{1}{2}\partial_j\partial_i\gamma_{ww}
        -\gamma_{mn}\Gamma^n_{ij}\Gamma^m_{ww}
        \nonumber \\
  &     - \frac{1}{2}\frac{\partial_z \gamma_{ij}}{z}
        +\frac{\delta_{z(i}\gamma_{j)z}
        -\delta_{iz}\delta_{jz}\gamma_{ww}}{z^2}
        +\frac{1}{4}\gamma^{ww}\partial_i\gamma_{ww}\partial_j\gamma_{ww},
        \label{eq:R3iwjw} \\
  \Gamma^m_{ww}\equiv & -\frac{1}{2}\gamma^{ml}\partial_l\gamma_{ww}
        +\frac{\delta^m_z-\gamma^{mz}\gamma_{ww}}{z},
        \label{eq:Gamma3mww} \\
  \mathcal{R}_{abcl} = & 0, \\
  \mathcal{R}_{abcd} = & (\delta_{ac}\delta_{bd}
        -\delta_{bc}\delta_{ad})\mathcal{R}_{wuwu}, \\
  \mathcal{R}_{wuwu} \equiv & -\frac{1}{4} \gamma^{mn}
        \partial_m\gamma_{ww}\partial_n\gamma_{ww}
        -\gamma_{ww}\frac{\gamma^{zm}}{z}\partial_m\gamma_{ww}
        +\frac{\gamma_{ww}-\gamma^{zz}\gamma_{ww}^2}{z^2}.
        \label{eq:R3wuwu}
\end{align}
For the right-hand side of Eq.~(\ref{contrGauss}), we also need
the spatial Ricci tensor which
is obtained from contraction of the Riemann tensor over the
first and third index. In $SO(D-d)$ symmetry, its non-vanishing components are
\begin{align}
  \mathcal{R}_{ij} = & \gamma^{mn}\mathcal{R}_{minj}
        +(D-d-1)\gamma^{ww}\mathcal{R}_{iwjw}\,, \\
  \mathcal{R}_{ab} = & \delta_{ab}\mathcal{R}_{ww}\,, \\
  \mathcal{R}_{ww} \equiv & \gamma^{mn}\mathcal{R}_{mwnw}
        + (D-d-2)\gamma^{ww}\mathcal{R}_{wuwu} \,.
\end{align}
Note that the last expression, $\gamma^{ww}\mathcal{R}_{wuwu}$,
does {\em not} involve a summation over $w$, but merely stands for
the product of $\gamma^{ww}$ with the expression (\ref{eq:R3wuwu}).

The components of the extrinsic curvature are given by Eq.~(\ref{eq:ADMSO3}).
Its derivative is directly obtained from the expressions (A.1)-(A.12)
in Appendix A of \cite{Cook:2016soy} and can be written as
\begin{align}
  D_iK_{jk} = & \partial_iK_{jk}-\Gamma^l_{ij}K_{kl}-\Gamma^l_{ik}K_{lj}\,, \\
  D_iK_{ab} = & \delta_{ab}(\partial_iK_{ww}-K_{ww}\gamma^{ww}
        \partial_i\gamma_{ww})\,, \\
  D_aK_{bj} = & \delta_{ab}\left(\frac{K_{jz}-\delta_{jz}K_{ww}}{z}
        -\frac{1}{2}K_{ww}\gamma^{ww}\partial_j\gamma_{ww}
        -K_{ij}\Gamma^i_{ww}\right)\,.
  \label{eq:EK3}
\end{align}
Next, we plug the expressions assembled in
Eqs.~(\ref{eq:spatialRiemann})-(\ref{eq:EK3}) into
the Gauss-Codazzi equations (\ref{Gauss})-(\ref{contrGauss})
where, we recall, early Latin indices $A,\,B,\,\ldots$ are now
replaced by $I,\,J,\,\ldots$ following our switch to
adapted coordinates. Splitting the index range $I$ into
$(i,\,a)$ for components inside and outside the computational domain,
and recalling that an index 0 denotes contraction with $\boldsymbol{n}$,
we can write the resulting components of the spacetime Riemann tensor as
\begin{align}
  R_{ijkl} & =  \mathcal{R}_{ijkl} + K_{ik}K_{jl} - K_{il} K_{jk},
        \label{eq:Rijkl} \\
  R_{ibkd} & =  \delta_{bd}R_{iwkw} \label{eq:Ribkd},\\
  R_{iwkw} & \equiv  \mathcal{R}_{iwkw} + K_{ik} K_{ww},\\
  R_{abcd} & =  (\delta_{ac}\delta_{bd}-\delta_{bc}\delta_{ad})
        (\mathcal{R}_{wuwu}+K_{ww}^2), \label{eq:Rabcd}\\
  R_{ajkl} & = R_{abcl}=0, \\
  R_{i0kl} & =  D_lK_{ik}-D_kK_{il}, \\
  R_{a0ck} & =  \delta_{ac}R_{w0wk}, \label{eq:Ra0ck}\\
  R_{w0wk} & \equiv  \partial_k K_{ww}-\frac{1}{2} \gamma^{ww} K_{ww}
        \partial_k\gamma_{ww} - \frac{K_{kz}-\delta_{kz}K_{ww}}{z}
        +\Gamma^m_{ww}K_{mk},
        \label{eq:Rw0wk} \\
  R_{a0cd} & =  R_{i0kd} = R_{a0kl} = 0, \\
  R_{i0j0} & =  \mathcal{R}_{ij}+KK_{ij}-K_{im}K^m_j, \\
  K & =  \gamma^{mn}K_{mn}+(D-d-1)\gamma^{ww}K_{ww}, \\
  R_{a0b0} & = \delta_{ab}R_{w0w0} \label{eq:Ra0b0},\\
  R_{w0w0} & \equiv  \mathcal{R}_{ww} + (K - \gamma^{ww}K_{ww})K_{ww}, \\
  R_{a0i0} & =  0. \label{eq:Ra0i0}
\end{align}
With these expressions, we are able to calculate all components of
the spacetime Riemann tensor directly from the ADM variables
$\gamma_{ij}$, $\gamma_{ww}$, $K_{ij}$ and $K_{ww}$ and their spatial
derivatives. There remains, however, one subtlety arising from the
presence of terms containing explicit division by $z$. Numerical
codes employing vertex centered grids need to evaluate these
terms at $z=0$. As described in detail in
\ref{sec:regularization}, all the above terms involving
division by $z$ are indeed regular and can be rewritten in a form
where this is manifest with no divisions by zero.


\subsection{The null Frame} \label{mvecs}
The null frame we need for the projections of the Weyl tensor
consists of $D$ unit vectors as given in Eq.~(\ref{eq:tetrad}): (i) the
ingoing null vector $k^{\A}$, (ii) the outgoing null vector $l^A$ which,
however, does not explicitly appear in the scalars (\ref{eq:weylcomps})
for the outgoing gravitational radiation, and (iii) the $(D-2)$ vectors
$m_{(\alpha)}^A$ pointing in the angular directions on the sphere.

We begin this construction with the $D-2$ unit basis vectors on the
$D-2$ sphere, $m^{\A}_{(\alpha)}$, and recall for this purpose
Eq.~(\ref{eq:coordinates}) that relates our
spherical coordinates $(r,\phi^{\alpha})$ to the Cartesian
$(x^{\hat{i}},\,z,\,w^a)$.
The set of spatial vectors, although not yet in orthonormalised form,
then consists of a radial vector denoted by $\tilde{m}_{(1)}$
and $D-2$ angular vectors $\tilde{m}_{(\alpha)}$ whose components
in Cartesian coordinates $X^{\I}=(x^{\hat{i}},\,z,\,w^a)$
on the computational domain $w^a=0$
are obtained through chain rule
\begin{align}
  \tilde{m}_{(1)} &= \frac{\partial}{\partial r}
        = \frac{\partial X^{\I}}{\partial r}
          \frac{\partial}{\partial  X^{\I}}
        ~~~\Rightarrow~~~
  \tilde{m}_{(1)}^{\I}=\frac{1}{r}
        (x^1,\,\ldots,\,x^{d-1},\,z,\,0,\,\ldots,\,0)\,, \label{eq:radial}\\
  \tilde{m}_{(\alpha)} &= \frac{\partial}{\partial \phi^{\alpha}}
        = \frac{\partial X^{\I}}{\partial \phi^{\alpha}}
        \frac{\partial}{\partial X^{\I}}\,,
  \label{eq:cart2sphere}
\end{align}
We can ignore time
components here, because our coordinates are adapted to the space-time split,
so that all spatial vectors have vanishing time components and this feature
is preserved under the eventual Gram-Schmidt orthonormalisation.
Plugging Eq.~(\ref{eq:coordinates}) into (\ref{eq:cart2sphere}), we obtain for
$\tilde{m}_{(\alpha)}$ (after rescaling by
$r \times \sin \phi^2\times \ldots \times \sin \phi^{\alpha}$)
\begin{equation}
\underbrace{
  \left(
    \begin{matrix}
      -{\displaystyle \sum_{s=2}^{D-1} (w^{s})^2} \\[15pt]
      w^{1} w^{2} \\
      \vdots \\
      \vdots \\
      \vdots \\
      \vdots \\
      w^{1} w^{D-1}
    \end{matrix}
  \right)}_{=\tilde{m}^{\I}_{(2)}}
\,,\ldots\,,
\underbrace{
  \left(
    \begin{matrix}
      \left.
      \begin{matrix}
        0~ \\
        \vdots~ \\
        0~ \\
      \end{matrix}
      \right\} (\alpha-2) \times \\[15pt]
      - {\displaystyle \sum_{s=\alpha}^{D-1} (w^{s})^2} \\[15pt]
      w^{\alpha-1} w^{\alpha} \\
      \vdots \\
      w^{\alpha-1} w^{D-2} \\
      w^{\alpha-1} w^{D-1}
    \end{matrix}
  \right)}_{=\tilde{m}^{\I}_{(\alpha)}}
\,,\ldots\,,
\underbrace{
  \left(
    \begin{matrix}
      0 \\
      \vdots \\
      \vdots \\
      \vdots \\
      0 \\[5pt]
      -(w^{D-2})^2 - (w^{D-1})^2 \\[5pt]
      w^{D-3} w^{D-2} \\[5pt]
      w^{D-3} w^{D-1}
    \end{matrix}
  \right)}_{=\tilde{m}^{\I}_{(D-2)}}
\,, ~~
\underbrace{
  \left(
    \begin{matrix}
      0 \\
      \vdots \\
      \vdots \\
      \vdots \\
      0 \\[5pt]
      0 \\[5pt]
      -(w^{D-1})^2 \\[5pt]
      w^{D-2} w^{D-1}
    \end{matrix}
  \right)}_{=\tilde{m}^{\I}_{(D-1)}}
\,.
\label{eq:mtilde}
\end{equation}
In $D=4$ dimensions, these vectors, together with $\tilde{m}_{(1)}$ of
Eq.~(\ref{eq:radial})
would form the starting point
for Gram-Schmidt orthonormalisation; see e.g.~Appendix C in \cite{Sperhake:2006cy}.
In $D\ge 5$ dimensional spacetimes with $SO(D-d)$ symmetry,
however, we face an additional
difficulty: on the computational domain $w^a=0$, all components of
the vectors $\tilde{m}_{(d+1)},\,\ldots,\,\tilde{m}_{(D-1)}$ vanish
and their normalisation would result in divisions of zero by zero.
This difficulty is overcome by rewriting the Cartesian
components of the vectors in terms of spherical coordinates and then
exploiting the freedom we have in suitably orienting the frame. The details
of this procedure are given in \ref{coordtrans} where we derive
a manifestly regular set of spatial vectors given by
\begin{align}
  \tilde{m}_{(1)}^{\A} &= (0~|~x^1,\,\ldots,\,x^d~|~0,\ldots,\,0)\,,
        \label{eq:tildem1gen} \\
  \tilde{m}_{(2)}^{\A} &= (0~|~-\rho_2^2,~x^1x^2,~x^1x^3,~\ldots,~
        x^1 x^d~|~0,~\ldots,~0) \,,
        \label{eq:tildem2gen} \\
  \cdots & \cdots \nonumber\\
  \tilde{m}_{(\hat{\alpha})}^{\A} &= (0,~|~
        \underbrace{0,~\ldots,~0}_{(\hat{\alpha}-2)\times},~
        -\rho_{\hat{\alpha}}^2,~x^{\hat{\alpha}-1}x^{\hat{\alpha}},~
        \ldots,~
        x^{\hat{\alpha}-1}x^d~|~0,~\ldots,~0) \,,
        \label{eq:tildemigen} \\
  \cdots & \cdots \nonumber\\
  \tilde{m}_{(d)}^{\A} &= (0,~|~
        \underbrace{0,~\ldots,~0}_{(d-2)\times},~
        -x^d,~x^{d-1}~|~0,~\ldots,~0) \,,
        \label{eq:tildemdgen} \\
  \tilde{m}_{(d+1)}^{\A} &= (0~|~\underbrace{0,~\ldots,~0}_{d\times}~|~
        1,~0,~\ldots,~0) \,,
        \label{eq:tildemdp1gen} \\
  \cdots & \cdots \nonumber\\
  \tilde{m}_{(D-1)}^{\A} &= (0~|~\underbrace{0,~\ldots,~0}_{d\times}~|~
        0,~\ldots,~0,~1)\,,
        \label{eq:tildemDm1gen}
\end{align}
where $\rho_{\I}=\sum_{s=\I}^{D-1}(w^s)^2$, we have restored,
for completeness, the time component and the vertical
bars highlight the three component sectors: time, spatial on-domain,
and spatial off-domain. Equations
(\ref{eq:tildemdp1gen})-(\ref{eq:tildemDm1gen})
can, of course, be conveniently written in short-hand notation as
$\tilde{m}_{(a)}^A = \delta^A{}_{a}$.
For the special case $d=3$, the vectors are given by
\begin{align}
  \tilde{m}_{(1)}^{\A} &= (0~|~x,~y,~z~|~0,\ldots,~0)\,,
        \label{eq:tildem1} \\
  \tilde{m}_{(2)}^{\A} &= (0~|~-y^2-z^2,~xy,~xz~|~0,\ldots,~0) \,,
        \label{eq:tildem2} \\
  \tilde{m}_{(3)}^{\A} &= (0~|~0,~-z,~y~|~0,~\ldots,~0) \,,
        \label{eq:tildem3} \\
  \tilde{m}_{(4)}^{\A} &= (0~|~0,~0,~0~|~1,~0,~\ldots,~0) \,,
        \label{eq:tildem4} \\
  \cdots & \cdots \\
  \tilde{m}_{(D-1)}^{\A} &= (0~|~0,~0,~0~|~0,~\ldots,~0,~1)\,,
        \label{eq:tildemDm1}
\end{align}

The next step is to orthonormalise these vectors. Clearly the
vectors $m_{(a)}^A$ with components in the $w^a$ dimensions are
normalised by:
\begin{equation}
  m_{(a)}^A=\frac{1}{\sqrt{\gamma_{ww}}}\delta^{\A}{}_{a}
\end{equation}
For the remaining
$d$ vectors given by Eqs.~(\ref{eq:tildem1gen})-(\ref{eq:tildemdgen})
or, for $d=3$, the spatial triad consisting
of the three vectors
(\ref{eq:tildem1})-(\ref{eq:tildem3}),
we use standard Gram-Schmidt orthonormalisation.
Note that under this procedure
the components outside the computational domain of these
vectors remain zero and can therefore be ignored.

The final element of the null frame we need is the ingoing null
vector, which we call $k^{\A}$. Given in \cite{Godazgar:2012zq}
as $\partial/\partial u-\frac{1}{2}\partial / \partial \textswab{r}$
asymptotically, we transform out of Bondi coordinates, sending
$(u,\textswab{r}) \rightarrow (t,r)$ and furthermore use the
freedom of rescaling this null vector by applying a constant
factor of\footnote{The convention we adopt here is
more common (though not unanimous) in numerical relativity.} $\sqrt{2}$
\begin{equation}
  k^{\A}=\frac{1}{\sqrt{2}}\left( n^{\A}-m_{(1)}^{\A}\right)
\end{equation}
Expressing the timelike unit normal field
$n^{\A}$ in terms of our gauge
variables $\alpha, \beta^{\I}$ we find
\begin{equation}
  k^{\A}=\frac{1}{\sqrt{2}}\left(\frac{1}{\alpha},
        -\frac{\beta^{\I}}{\alpha}-m_{(1)}^{\I} \right),
  \label{eq:kA}
\end{equation}
where $\beta^{\I}=(\beta^i,\,0,\,\ldots,\,0),~m_{(1)}^{\I}=(m_{(1)}^i,\,0,
\ldots,\,0)$.
This result provides the ingoing null vector for any choice of $d$ and
is the version implemented in the code.

\subsection{The projections of the Weyl tensor}

Finally, we calculate the projections of the Weyl tensor that encode the
outgoing gravitational radiation
\begin{equation}
  \Omega'_{(\alpha)(\beta)}=R_{\A\B\CC\D}k^{\A}
        m_{(\alpha)}^{\B}k^{\CC}m_{(\beta)}^{\D}\,,
\end{equation}
[cf.~Eq.~(\ref{eq:weylcomps})] where $k^{\A}$ is given by
Eq.~(\ref{eq:kA}) and the normal frame vectors
$m_{(2)},\,\ldots,\,m_{(D-1)}$ are
those obtained from Gram-Schmidt orthonormalising
the right-hand sides of Eqs.~(\ref{eq:tildem1gen})-(\ref{eq:tildemDm1gen}).

We first note that $\Omega'_{(\alpha)(\beta)}$
is symmetric in $\alpha \leftrightarrow \beta$,
so contractions solely with $m_{(2)},\,\ldots,\,m_{(d)}$ will result in
$d(d-1)/2$ components $\Omega'_{(\hat{\alpha})(\hat{\beta})}$.
For the special case $d=3$, we obtain the three components
$\Omega'_{(2)(2)},~\Omega'_{(2)(3)},~\Omega'_{(3)(3)}$. The null
vector $k$ has vanishing $w$ components and from
Eqs.~(\ref{eq:Rijkl})-(\ref{eq:Ra0i0}) we see that all
components of the Riemann tensor where an odd number of indices is
in the range $a,\,b,\,\ldots$ are zero. The only non-vanishing
terms involving the Riemann tensor with off-domain indices
$a,\,b,\,\ldots$, therefore, have either four such indices
or two and contain a Kronecker delta
$\delta_{ab}$; cf.~Eqs.~(\ref{eq:Ribkd}), (\ref{eq:Rabcd}), (\ref{eq:Ra0ck}),
(\ref{eq:Ra0b0}). In consequence, the mixed components
$\Omega'_{(\hat{\alpha})(a)}=0$ and the purely off-domain
components $\Omega'_{(a)(b)} \propto \delta_{ab}$. The list of
all non-vanishing components $\Omega'_{(\alpha)(\beta)}$
is then given by
\begin{align}
  \Omega'_{(\hat{\alpha})(\hat{\beta})} = \frac{1}{4} &\left[
        R_{0k0l}m_{(\hat{\alpha})}^k m_{(\hat{\beta})}^l
        - R_{mk0l}m_{(1)}^m m_{(\hat{\alpha})}^k m_{(\hat{\beta})}^l
        - R_{0kml}m_{(\hat{\alpha})}^k m_{(1)}^m m_{(\hat{\beta})}^l
        \right. \nonumber \\
     &  \left.
        + R_{mknl}m_{(1)}^m m_{(\hat{\alpha})}^k m_{(1)}^n m_{(\hat{\beta})}^l
        \right]\,, \label{eq:Omegaij} \\[10pt]
  \Omega'_{(a)(b)} &= \delta_{ab} \, \Omega'_{(w)(w)}\,,\\[10pt]
  \Omega'_{(w)(w)} &= \frac{1}{4\gamma_{ww}}\left[R_{w0w0}
        - R_{w0wk}m_{(1)}^k
        - R_{w0wl}m_{(1)}^l
        + R_{wkwl}m_{(1)}^k m_{(1)}^l\right]\,.
  \label{eq:Omegaww}
\end{align}
where $\hat{\alpha},\hat{\beta}=2,\,\ldots,\,d$
and all components of the Riemann tensor
on the right-hand sides are listed in the set of
Eqs.~(\ref{eq:Rijkl})-(\ref{eq:Ra0i0}).
It should be noted here that $\Omega'_{(\alpha)(\beta)}$ is trace
free, and so $\Omega'_{(w)(w)}$ can be calculated from the diagonal
terms $\Omega'_{(2)(2)}, \ldots \Omega'_{(d)(d)}$.
In a numerical simulation, the components of $\Omega'_{(\alpha)(\beta)}$
are calculated as functions of time and then can be integrated
according to Eq.~(\ref{Mdot}) to extract the amount of energy
radiated in gravitational waves.

\subsection{$SO(2)$ symmetry}
In the axisymmetric case $d=D-2$
there exists only one $w$ direction (off domain).
As discussed in Section 4 of \cite{Cook:2016soy}, we
keep all tensor components as we would in the absence of symmetry,
and the modified Cartoon method and, thus, the rotational
symmetry only enters in the calculation
of spatial derivatives in the $w$ direction.
For $SO(2)$ symmetry, the extraction of gravitational waves therefore
proceeds as follows.
\begin{itemize}
  \item All components of the ADM metric and extrinsic curvature
        are extracted on the $D-2$ dimensional computational domain.
  \item The spatial Riemann tensor and its contractions are
        directly evaluated using Eq.~(\ref{eq:spatialRiemann})
        using the relations of Appendix C in \cite{Cook:2016soy}
        for off-domain derivatives.
  \item The necessary components of the spacetime Riemann tensor
        and its projections onto the timelike unit normal are
        evaluated through Eqs.~(\ref{Gauss})-(\ref{contrGauss}).
  \item The null frame is constructed as detailed in Sec.~\ref{mvecs},
        simply setting $d=D-2$.
  \item All the projections of the Weyl tensor onto the null frame vectors
        are obtained from Eq.~(\ref{eq:Omegaij}), but now covering the
        entire range of spatial indices
\begin{align}
  \Omega'_{(\alpha)(\beta)} = \frac{1}{4} &\left[
        R_{0\K 0\LL}m_{(\alpha)}^{\K} m_{(\beta)}^l
        - R_{\M \K 0\LL}m_{(1)}^{\M} m_{(\alpha)}^{\K} m_{(\beta)}^{\LL}
        - R_{0\K \M \LL}m_{(\alpha)}^{\K} m_{(1)}^{\M} m_{(\beta)}^{\LL}
        \right. \nonumber \\
     &  \left.
        + R_{\M \K \N \LL}m_{(1)}^{\M} m_{(\alpha)}^{\K}
        m_{(1)}^{\N} m_{(\beta)}^{\LL}
        \right]\,.
\end{align}
\end{itemize}
Note that with the existence of more components of the Riemann tensor,
more projections of the Weyl tensor now exist, specifically
cross-terms such as $\Omega'_{(2)(w)}$. This can be seen straightforwardly
by using $SO(2)$ modified Cartoon terms from appendix C of
\cite{Cook:2016soy} and the expressions for the full and spatial Riemann
tensor given in Eqs.~(\ref{Gauss}) and (\ref{eq:spatialRiemann}).
For example, we can
see that a component such as $R_{wijk}$ is non-zero. This will
contribute to terms of the form $\Omega'_{(\hat{\alpha})(w)}$. As already
emphasized in \cite{Cook:2016soy}, the key gain in employing the modified
Cartoon method for simulating axisymmetric spacetimes does {\em not}
lie in the elimination of tensor components, but in the
dimensional reduction of the computational domain.

\section{Numerical simulations}
\label{sec:Numerical}

In the remainder of this work, we will implement the specific version
of the wave extraction for the case $d=3$ and $D=6$ and
simulate head-on collisions of equal-mass, non-spinning BHs
starting from rest. We will calibrate
the numerical uncertainties arising from the numerical discretisation
of the equations
(fourth order in space and time and second order at the outer and
refinement boundaries),
the use of large but finite extraction radii and
also consider the dependency of the results on the initial separation
of the BHs. This type of collisions has already been studied
by Witek {\em et al.} \cite{Witek:2014mha} who calculate the GW
energy using the Kodama-Ishibashi formalism, which enables us to
compare our findings with their values.

\subsection{Code infrastructure and numerical Set-up}

We perform evolutions using the LEAN code
\cite{Sperhake:2006cy,Sperhake:2007gu} which is based on CACTUS
\cite{Cactusweb,Allen:1999}
and uses CARPET \cite{Carpetweb,Schnetter:2003rb}
for mesh refinement. The Einstein equations are implemented in the
BSSN formulation
with the modified Cartoon method employed to reduce computational
cost. For the explicit equations under the $SO(D-3)$ symmetry that
we use, see Section 3.2 of \cite{Cook:2016soy} with parameter $d=3$.
Without loss of generality, we perform collisions along the $x$-axis,
such that the centre-of mass is located at the origin of the
grid, and impose octant symmetry.

We specify the gauge in terms of the ``1+log'' and ``$\Gamma$ driver''
conditions for the lapse function and shift vector
(see e.g.~\cite{vanMeter:2006vi}) according to
\begin{align}
  \partial_t \alpha =& \beta^m \partial_m \alpha
  - 2\alpha K\,, \\
  \partial_t \beta^i =& \beta^m \partial_m \beta^i
  + \frac{1}{4} \tilde{\Gamma}^i - \frac{1}{2^{1/3}R_h} \beta^i\,,
\end{align}
with initial values $\alpha=1$, $\beta^i=0$.

The BH initial data is calculated
using the higher dimensional generalization
of Brill-Lindquist data \cite{Brill:1963yv,Zilhao:2011yc}
given in terms of the ADM variables by
\begin{equation}
  K_{\I \J} = 0\,,~~~
  \gamma_{\I \J} = \psi^{4/(D-3)} \delta_{\I \J}\,,~~~
  \psi = 1 + \sum_{\mathcal{N}} \frac{\mu_{\mathcal{N}}}
  {4 \left[\sum_{\K=1}^{D-1}
  (X^{\K} - X_{\mathcal{N}}^{\K})^2\right]^{(D-3)/2}} \,,\label{BLid}
\end{equation}
where the summation over $\mathcal{N}$ and $K$ extends over the multiple
BHs and spatial coordinates, respectively, and $X^{\K}_{\mathcal{N}}$
denotes the position of the $\mathcal{N}^{\rm th}$ BH. As mentioned above,
we place the BHs on the $x$ axis in the centre-of-mass frame, so
that in the equal-mass case, we have $X^1=\pm x_0$. Our initial
configuration is therefore completely specified by
the initial separation which we measure in units of the horizon
radius $R_h$ of a single BH. The BH mass
and the radius $R_h$ are related through the mass parameter $\mu$
by
\begin{equation}
  \mu = \frac{16\pi M}{(D-2)\mathcal{A}_{D-2}}\,,~~~~~
  \mu = R_{\rm h}^{D-3}\,,~~~~~
  \mathcal{A}_{D-2}=\frac{2\pi^{(D-1)/2}}{\Gamma\left(\frac{D-1}{2}\right)}\,,
  \label{eq:RhMmu}
\end{equation}
where $\mathcal{A}_{D-2}$ is the surface area of the unit $(D-2)$ sphere.

The computational domain used for these simulations consists of a set
of eight nested refinement levels which we characterize in terms of the
following parameters: (i) the resolution $h$ on the innermost level
which gets coarser by a factor of two on each consecutive outer level,
(ii) the size $L$ of the domain which describes the distance
of the outermost edge from the origin, and (iii) the resolution
$H$ on the refinement level where the gravitational waves are extracted.

For each simulation,
we calculate the $\Omega'_{(\alpha)(\beta)}$
on our three dimensional computational
grid and project them onto a two dimensional array representing a spherical
grid at fixed coordinate radius.
The data thus obtained on the extraction sphere are inserted
into Eq.~(\ref{Mdot}).
The $\Omega'_{(\alpha)(\beta)}$ are scalars and so in our angular coordinate
system do not depend on $\phi^{4},\ldots,\phi^{D-1}$, so the integral
over the sphere in (\ref{Mdot}) can be simplified:
\begin{equation}
  \dot{M}(u)=-\lim_{r\rightarrow \infty} \frac{r^{D-2}}{8\pi}
        \mathcal{A}_{D-4} \int^\pi_0\int^\pi_0I[{\Omega'}^2]~\sin^{D-3}(\phi^2)
        ~\sin^{D-4}(\phi^3)~d\phi^3d\phi^2\,,
\end{equation}
where $I[\Omega'^2]\equiv \left( \int_{-\infty}^{u} \Omega'_{(\alpha)(\beta)}
d\tilde{u}\right)^2$.
A final integration over time of the variable $\dot{M}$ then gives the
total radiated energy.

\subsection{Numerical Results} \label{sec:Results}

We begin our numerical study with an estimate of the uncertainty
in our GW estimates arising from the discretisation of the equations.
For this purpose, we have evolved two BHs initially located at
at $x=\pm x_0=\pm 4.0~R_h$ using a computational grid of size
$L=181~R_h$ and three resolutions $h_1=R_h/50.8$, $h_2=R_h/63.5$
and $h_3=R_h/76.2$ which corresponds to $H_1=R_h/2.12$, $H_2=R_h/2.65$
and $H_3=R_h/3.17$ in the extraction zone.

\begin{figure}
\centering
\includegraphics[width=0.8\textwidth]{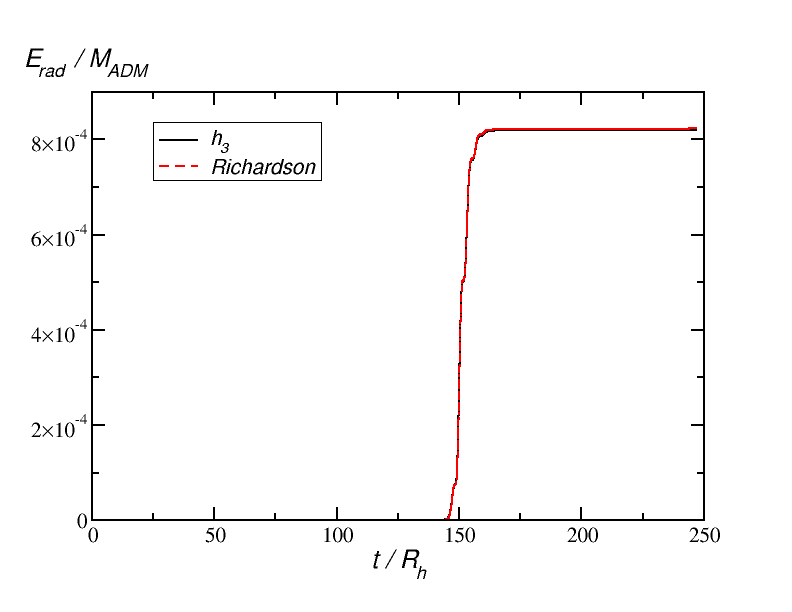}
\includegraphics[width=0.8\textwidth]{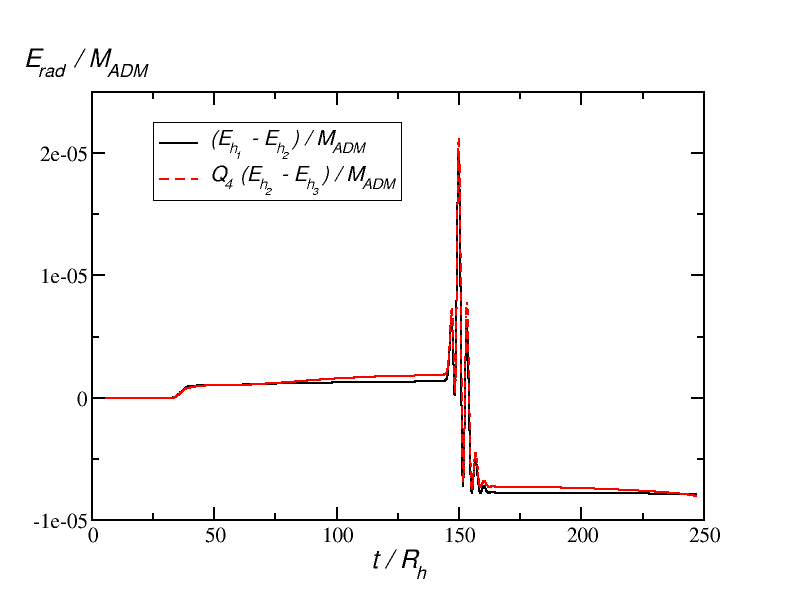}
\caption{Upper panel: Radiated energy as a function of time obtained for
         the highest resolution $h_3=R_h/76.2$ (solid curve) and Richardson
         extrapolated to infinite resolution assuming fourth-order
         convergence (dashed curve).
         Lower panel: Convergence plot for the radiated energy
         $E_{\rm rad}$ extracted at $r_{\rm ex}=50.4~R_h$
         from an equal-mass collision of two non-spinning
         BHs in $D=6$ starting from
         a separation $8~R_h$. The results shown have been obtained
         using resolutions $h_1=R_h/50.8$, $h_2=R_h/63.5$
         and $h_3=R_h/76.2$. The difference in radiated energy between
         the medium and high-resolution simulations has been rescaled
         by a factor $Q_4=2.784$
         expected for fourth-order convergence.}
  \label{fig:convergence}
\end{figure}

We measure the radiated energy in units of the total ADM mass
of the spacetime, which for Brill-Lindquist data
is given by Eq.~(\ref{eq:RhMmu}) with
$\mu \equiv \mu_1 + \mu_2$, the mass parameters of the initial BHs.
The radiated energy as a function of time is shown in
the upper panel of Fig.~\ref{fig:convergence}. The radiation
is almost exclusively concentrated within a window of $\Delta t \approx
20~R_h$ around
merger. During the infall
and the post-merger period, in contrast, $E_{\rm rad}$ remains
nearly constant. By comparing the high-resolution result
with that obtained for the coarser grids, we can test the order of
convergence. To leading order, the numerical result $f_{h}$ for some
variable obtained at finite resolution $h$ is related to the continuum
limit solution $f$ by $f=f_h + \mathcal{O}(h^n)$, where $n$ denotes
the order of convergence. By evaluating the quotient
\begin{equation}
  Q_n = \frac{f_{h_1}-f_{h_2}}{f_{h_2}-f_{h_3}}
        = \frac{(h_1/h_2)^n-1}{1-(h_3/h_2)^n}\,,
\end{equation}
we can then plot the two differences $f_{h_1}-f_{h_2}$ and
$f_{h_2}-f_{h_3}$ and test whether their ratio is consistent with
a given value $n$.
The results for our study are shown in the lower panel
of Fig.~\ref{fig:convergence} which demonstrates that
our numerical results converge at fourth order.
The discretisation error of the total radiated energy is then obtained
as the difference between the finite resolution result and that
predicted by Richardson extrapolation (see upper panel in the figure).
We obtain for the high-resolution case
$E_{\rm rad}=8.19\times 10^{-4}~M_{\rm ADM}$
with a discretisation error of $\sim 0.4~\%$.

A second source of error arises from the extraction at finite radius.
Following standard practice (see e.g.~\cite{Hinder:2013oqa}),
we estimate this uncertainty by extracting the GW energy at
a set of seven or eight finite radii in the range $40~R_h$ to $110~R_h$
and extrapolating these values assuming a
functional dependency
\begin{equation}
  E_{\rm rad}(r) = E_{\rm rad}(\infty) + \frac{a}{r}
        + \mathcal{O}\left( \frac{1}{r^2} \right)\,,
\end{equation}
where $a$ is a coefficient obtained through the fitting of the numerical
data. By applying this procedure, we estimate the uncertainty due to
the extraction radius at $0.2~\%$ at $R_{\rm ex}=110~{\rm R_h}$ and
$0.4~\%$ at $R_{\rm ex}=60~R_h$.

Finally, we have measured the dependency of the total radiated energy
on the initial separation of the BHs. In addition to the simulations
discussed so far, we have performed high-resolution
simulations placing the BHs at $x_0=\pm 7.8~R_h$ and at
$\pm 12.8~R_h$. We have found very small variations at a level
of $0.1~\%$ in the radiated energy for these cases, well below
the combined error budget of $0.6~\%$ obtained above.
Compared with collisions in $D=4$ dimensions
(see e.g.~Table II in \cite{Sperhake:2006cy}), $E_{\rm rad}$ shows significantly
weaker variation with initial separation in $D=6$.
We attribute this to the more rapid fall-off of the force
of gravity in higher dimensions leading to a prolonged but dynamically
slow infall phase which generates barely any GWs.

In summary, we find the total energy radiated in gravitational waves
in a head-on collision of two equal-mass, non-spinning BHs to be
\begin{equation}
  E_{\rm rad} = (8.19\pm 0.05) \times 10^{-4}~M_{\rm ADM}\,,
\end{equation}
in excellent agreement with the value
$(8.1\pm 0.4)\times 10^{-4}$ reported in \cite{Witek:2014mha}
using the Kodama-Ishibashi formalism.

\section{Conclusions}
The extraction of gravitational waves from numerical simulations is one
of the most important diagnostic tools in studying the strong-field
dynamics of compact objects in four as well as higher dimensional
spacetimes. In this work we have formulated the Weyl tensor based
wave extraction technique of Godazgar \& Reall \cite{Godazgar:2012zq}
-- a higher dimensional generalization of the Newman-Penrose scalars --
in a form suitable for numerical simulations of $D>4$ dimensional
spacetimes with $SO(D-d)$, $1\le d\le D-2$, symmetry employing the
modified Cartoon method. The only prerequisite for implementing
our formalism is the availability on each timelike hypersurface
of the effective computational domain of the ADM variables. These
are constructed straightforwardly from all commonly used
numerical evolution systems such as BSSN, generalized harmonic or
conformal Z4.

The recipe for extracting the GW signal then consists of the
following steps.
\begin{list}{\rm{{\bf(\arabic{count})}}}{\usecounter{count}
             \labelwidth0.5cm \leftmargin0.7cm \labelsep0.2cm \rightmargin0cm
             \parsep0.5ex plus0.2ex minus0.1ex \itemsep0ex plus0.2ex}
\item
  Computation of the on and off-domain components of the spatial
  Riemann tensor (which equals the Weyl tensor in the vacuum
  extraction region) and the derivative of the extrinsic curvature
  according to Eqs.~(\ref{eq:R3ijkl})-(\ref{eq:EK3}).
\item
  Reconstruction of the components of the spacetime Riemann
  tensor as well as its contractions with the unit timelike normal
  from the quantities of the previous step according to
  Eqs.~(\ref{eq:Rijkl})-(\ref{eq:Ra0i0}).
\item
  Construction of the null-frame vectors through Gram-Schmidt orthonormalising
  the expressions of Eqs.~(\ref{eq:tildem1gen})-(\ref{eq:tildemDm1gen})
  and then using (\ref{eq:kA}) for the ingoing null vector.
\item
  Calculation of the projections $\Omega'_{(\alpha)(\beta)}$
  of the Weyl tensor onto the null frame vectors
  using Eqs.~(\ref{eq:Omegaij})-(\ref{eq:Omegaww}).
\item
  Calculation of the energy flux in GWs through Eq.~(\ref{Mdot})
  and integration in time of the flux to obtain the total
  radiated energy.
\end{list}
The most common case of modeling higher dimensional spacetimes
with rotational symmetries is the case of $d=3$ effective spatial
dimensions which allows for straightforward generalization
of existing codes (typically developed for 3+1 spacetimes)
and also accommodates
sufficiently complex dynamics to cover most of the important
applications of higher dimensional numerical relativity.
We have, for this purpose, explicitly given the specific
expressions of some of our relations for $d=3$ where these
are not trivially derived from their general counterparts.

For testing the efficacy and accuracy of this method, we have
applied the wave extraction to the study of equal-mass, non-spinning
headon collisions of BHs starting from rest in $D=6$ using
$d=3$. We find these collisions to radiate a fraction
$(8.19 \pm 0.05) \times 10^{-4}$ of the ADM mass in GWs,
in excellent agreement with a previous study \cite{Witek:2014mha}
employing a perturbative extraction technique based on the
Kodama-Ishibashi formalism. We find this energy to be essentially
independent of the initial separation which we have varied from
$8.0$ to $15.6$ and $25.6$ times the horizon radius of a single
BH. We attribute this result to the higher fall-off rate of the
gravitational attraction in higher dimensions and the correspondingly
slow dynamics during the infall stage.

We finally note that the Weyl tensor based wave extraction ideally
complements the perturbative extraction technique of the Kodama-Ishibashi
formalism. The latter provides the energy contained in individual
$(l,m)$ radiation multipoles but inevitably requires cutoff at
some finite $l$. In contrast, the $\Omega'_{(\alpha)(\beta)}$
facilitate calculation of the total radiation, but without
multipolar decomposition. It is by putting both extraction techniques
together, that we obtain a comprehensive description of the entire
wave signal. Future applications include the stability of highly
spinning BHs and their transition from unstable to stable
configurations, the wave emission in evolutions of black rings
and an extended study of higher dimensional BH collisions over a
wider range of dimensionality $D$, initial boosts and with non-zero
impact parameter.
These studies require particularly high resolution to accurately model
the rapid fall-off of gravity, especially for $D\gg 4$, and are therefore
beyond the scope of the present study. However, the foundation for
analysing in detail the GW emission in these and many more scenarios
is now available in as convenient a form as in the more traditional
3+1 explorations of numerical relativity.

\section*{Acknowledgments}
We thank Pau Figueras, Mahdi Godazgar, Markus Kunesch, Harvey Reall, Saran Tunyasuvunakool and Helvi Witek for highly
fruitful discussions of this topic.
This work has received funding from the European Union’s Horizon
2020 research and innovation programme under the Marie Sk\l odowska-Curie
grant agreement No 690904, from H2020-ERC-2014-CoG Grant No.~”MaGRaTh"
646597,
from STFC Consolidator Grant No. ST/L000636/1,
the SDSC Comet and TACC Stampede clusters through NSF-XSEDE Award
Nos.~PHY-090003,
the Cambridge High Performance
Computing Service Supercomputer Darwin using Strategic Research
Infrastructure Funding from the HEFCE and the STFC, and DiRAC's Cosmos
Shared Memory system through BIS Grant No.~ST/J005673/1 and STFC Grant
Nos.~ST/H008586/1, ST/K00333X/1.
W.G.C. is supported by a STFC studentship.

\appendix
\section{Regularisation of terms at $z=0$}
\label{sec:regularization}
\paragraph{}
For the axisymmetric case $d=D-2$, we only need to regularise
terms appearing in the calculation of derivatives in the off-domain
$w$ direction. All these terms are given explicitly in
Appendix C of \cite{Cook:2016soy}, so that in the following
we can focus exclusively on the additional terms appearing for
$1\le d \le D-3$, i.e.~for spacetimes admitting two or more rotational
Killing vector fields.

The treatment of these terms proceeds in close analogy to that of the
BSSN equations in the modified Cartoon approach as described
in detail in Appendix B of \cite{Cook:2016soy}.
In contrast to
that work, however,
we will not
be using the conformally rescaled metric of the BSSN equations,
which satisfies the simplifying condition $\det\tilde{\gamma}=1$,
and so certain regularised terms involving the inversion of the
metric will differ from the expressions obtained for the
BSSN equations.

We start with a brief summary of the techniques and the main
assumptions we will use to regularise expressions:

\vspace{0.3cm}
\noindent
{\bf \em 1.~Regularity:} We require all tensor components and their derivatives
to be regular when
expressed in Cartesian coordinates. Under transformation
to spherical coordinates this implies that tensors containing
an odd (even) number of radial indices, i.e.~$z$ indices in
our notation, contain exclusively odd (even) powers of $z$
in a series expansion around $z=0$. Using such a series expansion
enables us to trade divisions by $z$ for derivatives with respect to $z$.
For example, for the $z$ component of a vector field $\boldsymbol{V}$,
we obtain
\begin{align}
  \frac{V^z}{z}= & \frac{a_1 z+a_3 z^3 +\ldots}{z}= a_1+a_3z^2+\ldots
  \overset{*}{=}
  a_1 \overset{*}{=} \partial_z V^z
  \,,
\end{align}
where we have introduced the symbol $\overset{*}{=}$ to denote equality
in the limit $z\rightarrow 0$.

\vspace{0.3cm}
\noindent
{\bf \em 2.~Absence of conical singularities:} We require that the
spacetime contain no conical singularity
at the origin $z=0$. For the implications of this condition, we
consider the coordinate transformation from $(x^{\hat{i}},z,w^{d+1}, \ldots,
w^a\ldots, w^{D-1})$ to $(x^{\hat{i}},\rho, w^{d+1},\ldots,w^{a-1}, \varphi,
w^{a+1},\ldots, w^{D-1})$.
As no other $w^b,~b\ne a$, coordinates will enter
into this discussion we shall refer to $w^a$ as $w$.
In these coordinates we have that
\begin{align}
  \gamma_{\rho\rho} = & \frac{z^2}{\rho^2}\gamma_{zz}
      +2\frac{zw}{\rho^2}\gamma_{zw}+\frac{w^2}{\rho^2}\,, \\
  \gamma_{\varphi \varphi} = & w^2\gamma_{zz}-2wz\gamma_{zw}+z^2\gamma_{ww}\,,
\end{align}
and the line element for vanishing $dx^{\hat{i}}=0$ and $dw^b = 0$,
$b\ne a$, is given by
\begin{align}
  ds^2= & \gamma_{\rho\rho} d\rho^2+\rho^2\gamma_{\varphi\varphi}
      d\varphi^2\,.
\end{align}
Requiring the circumference to be the radius
times $2\pi$, we have that
$\gamma_{\varphi\varphi}=\rho^2\gamma_{\rho\rho}$.
Substituting the above expressions and taking the limit $z
\rightarrow 0$, we obtain
\begin{align}
  \gamma_{zz}-\gamma_{ww} \overset{*}{=} \mathcal{O}(z^2)\,.
\end{align}
Taking the time derivative of this relation and using the definition
of the extrinsic curvature, we find that likewise
\begin{align}
  K_{zz}-K_{ww} \overset{*}{=} \mathcal{O}(z^2)\,.
\end{align}

\vspace{0.3cm}
\noindent
{\bf \em 3.~Inverse metric}: Various terms that we need to address
contain factors of
the inverse metric $\gamma^{\I\J}$. In the practical regularisation
procedure, these terms are conveniently handled by writing
expressing $\gamma^{\I\J}$ in terms of the downstairs metric
components $\gamma_{ij}$ and $\gamma_{ww}$ which are the fields
we evolve numerically.
We know the metric takes the following form:
\begin{equation}
  \gamma_{\I\J} =
  \left(
  \begin{array}{cccc|cccc}
    \gamma_{x^1x^1} & \cdots & \gamma_{x^1x^{d-1}} & \gamma_{x^1z} & 0 & 0 &
        \cdots & 0 \\
    \vdots & \ddots &\vdots &\vdots & \vdots & \vdots & \cdots & \vdots \\
    \gamma_{x^{d-1}x^1} & \cdots & \gamma_{x^{d-1}x^{d-1}} &
        \gamma_{x^{d-1}z} & 0 & 0 & \cdots & 0 \\
    \gamma_{zx^1} & \cdots & \gamma_{zx^{d-1}} & \gamma_{zz} & 0 & 0 &
        \cdots & 0 \\
    \hline
    0   & \cdots   & 0      & 0      & \gamma_{ww} & 0 & \ldots & 0 \\
    0   & \cdots   & 0      & 0      & 0 & \gamma_{ww} & \ldots & 0 \\
    \vdots & \cdots & \vdots & \vdots & \vdots              & \vdots &
        \ddots & \vdots \\
    0      & \cdots & 0      & 0      & 0                   & 0 & \cdots &
        \gamma_{ww}
  \end{array}
  \right)
  \label{eq:gammamatrix}
  \,,
\end{equation}
and we shall denote the upper left quadrant by the matrix $M_{ij}$.
For simplicity, we will use the index $\hat{i}$ to denote $x^{\hat{i}}$
in this section, so e.g.~cofactors $C_{12}=C_{x^1x^2}$ and
$C_{1z}=C_{x^1z}$, and similarly indices $i,\,j,\,...$ will denote the
same range, but including the $z$ component.

We can now denote the cofactor of an element in the top left quadrant
of $\gamma_{\I\J}$ as
\begin{equation}
  C_{ij}=(-1)^{i+j}\gamma_{ww}^\eta \det(M_{kl\{k\ne j, l\ne i\}})
  \label{eq:cofactord}
\end{equation}
where $\eta=D-d-1$ and
the notation $\det(M_{kl\{k\ne j, l\ne i\}})$ denotes the
determinant of the matrix $M_{kl}$ obtained by crossing out the $j^{\rm th}$
row and $i^{\rm th}$ column. Likewise, we may add further inequalities
inside the braces to denote matrices obtained by crossing out more
than one row and column. We can then insert this expression for
$C_{ij}$ and the determinant of the right hand side of
Eq.~(\ref{eq:gammamatrix}),
\begin{align}
  \det \gamma_{\I\J} &= \gamma_{ww}^{\eta} \det \gamma_{ij} \nonumber \\
        &\overset{*}{=} \gamma_{ww}^{\eta} \gamma_{zz}
        \det (M_{kl\{k\ne z,l\ne z\}})\,,
  \label{eq:detgIJzgen}
\end{align}
to obtain expressions for inverse metric components according to
\begin{align}
  \gamma^{ij}=\frac{C_{ij}}{\det \gamma_{\I\J}}\,.
  \label{eq:invg}
\end{align}
For $d=3$, this procedure starts from the spatial metric
\begin{align}
  \gamma_{\I\J} =
  \left(
  \begin{array}{ccc|ccc}
    \gamma_{xx} & \gamma_{xy} & \gamma_{xz} & 0 & \cdots & 0 \\
    \gamma_{yx} & \gamma_{yy} & \gamma_{yz} & 0 & \cdots & 0 \\
    \gamma_{zx} & \gamma_{zy} & \gamma_{zz} & 0 & \cdots & 0 \\
    \hline
    0      & 0      & 0      & \gamma_{ww} & \ldots & 0 \\
    \vdots & \vdots & \vdots & \vdots              & \ddots & \vdots \\
    0      & 0      & 0      & 0                   & \cdots & \gamma_{ww}
  \end{array}
  \right)\,.
\end{align}
The components $C_{ij}$ of the cofactor matrix (which is symmetric) are
given by
\begin{align}
{\footnotesize
  \begin{array}{lll}
    C_{xx} = \gamma_{ww}^n (\gamma_{yy} \gamma_{zz}
      - \gamma_{yz}^2)\,,~&
    C_{xy} = -\gamma_{ww}^n (\gamma_{yx} \gamma_{zz}
      - \gamma_{zx} \gamma_{yz})\,,~&
    C_{xz} = \gamma_{ww}^n ( \gamma_{yx} \gamma_{zy}
      - \gamma_{zx} \gamma_{yy})\,, \\[10pt]
    \cdots &
    C_{yy} = \gamma_{ww}^n (\gamma_{xx} \gamma_{zz}
      - \gamma_{zx}^2)\,, &
    C_{yz} = -\gamma_{ww}^n (\gamma_{xx} \gamma_{zy}
      - \gamma_{zx} \gamma_{xy})\,, \\[10pt]
    \cdots & \cdots &
    C_{zz} = \gamma_{ww}^n (\gamma_{xx} \gamma_{yy}
      - \gamma_{xy}^2)\,,
  \end{array}
}
\nonumber \\[10pt]
\label{eq:cofactor}
\end{align}
the determinant becomes
\begin{align}
  \det \gamma_{\I\J} &= \gamma_{ww}^{\eta} \left(
  \gamma_{xx} \gamma_{yy} \gamma_{zz}
  + 2 \gamma_{xy} \gamma_{xz} \gamma_{yz}
  - \gamma_{xx} \gamma_{yz}^2
  - \gamma_{yy} \gamma_{xz}^2
  - \gamma_{zz} \gamma_{xy}^2 \right)
  \nonumber \\
  &\overset{*}{=}
  \gamma_{ww}^{\eta} \gamma_{zz}
  \left(
  \gamma_{xx} \gamma_{yy} - \gamma_{xy}^2
  \right)\,,
  \label{eq:detgIJz}
\end{align}
and the inverse metric follows by inserting these into
Eq.~(\ref{eq:invg}).

\vspace{0.3cm}
\noindent
Using these techniques, we can regularise all terms in
Eqs.~(\ref{eq:R3iwjw}), (\ref{eq:Gamma3mww}), (\ref{eq:R3wuwu}),
(\ref{eq:EK3}) and (\ref{eq:Rw0wk}) that contain divisions by $z$.
It turns out to be convenient to combine the individual terms into the following six
expressions.

\vspace{0.3cm}
\noindent
\begin{list}{\rm{{\bf(\arabic{count})}}}{\usecounter{count}
             \labelwidth0.5cm \leftmargin0.7cm \labelsep0.2cm \rightmargin0cm
             \parsep0.5ex plus0.2ex minus0.1ex \itemsep0ex plus0.2ex}
\item
  $$\frac{\delta^i_z-\gamma^{zi}\gamma_{ww}}{z}$$\\
  We express $\gamma^{zi}$ in terms of the metric, and trade divisions
  by $z$ for derivatives $\partial_z$ and obtain
  \begin{align}
    \frac{\delta^i_z-\gamma^{zi}\gamma_{ww}}{z} \overset{*}{=}& \left\{
    \begin{array}{ll}\displaystyle
      \sum^{d-1}_{\hat{j}=1} (-1)^{\hat{i}+\hat{j}}
        \frac{\gamma_{ww}}{\det{M_{mn}}}\partial_z\gamma_{z\hat{j}}
        \det(M_{kl\{k\ne\hat{i}, k\ne z, l\ne z, l\ne \hat{j}\}})~~ &
      \text{if }i = \hat{i} \\[10pt]
      0 & \text{if }i = z
    \end{array}
    \right.\\ \nonumber
  \end{align}
  For the $d=3$ case this reduces to
  \begin{align}
    \frac{\delta^i_z-\gamma^{zi}\gamma_{ww}}{z} \overset{*}{=}& \left\{
    \begin{array}{ll}\displaystyle
      \frac{\gamma_{yy} \partial_z \gamma_{xz}
      - \gamma_{xy} \partial_z \gamma_{yz}}{\gamma_{xx}\gamma_{yy}
      -\gamma_{xy}^2}~~~~ &
      \text{if }i = x \\[10pt]
      \displaystyle\frac{\gamma_{xx} \partial_z \gamma_{yz}
      - \gamma_{xy} \partial_z \gamma_{xz}}{\gamma_{xx}\gamma_{yy}
      -\gamma_{xy}^2}~~~~ &
      \text{if }i = y \\[10pt]
      0 & \text{if }i = z
    \end{array}
    \right.\,.\\ \nonumber 
  \end{align}

\item
  $$\frac{\partial_i\gamma_{jz}-\delta_{jz}\partial_i\gamma_{ww}}{z}
        -\delta_{iz}\frac{\gamma_{jz}-\delta_{jz}\gamma_{ww}}{z^2}
        +\frac{\partial_j\gamma_{iz}-\delta_{iz}\partial_j\gamma_{ww}}{z}
        -\delta_{jz}\frac{\gamma_{iz}-\delta_{iz}\gamma_{ww}}{z^2}$$ \\
  Here we simply trade divisions by $z$ for $\pd_z$ and obtain
  \begin{align}
    &\frac{\partial_i\gamma_{jz}-\delta_{jz}\partial_i\gamma_{ww}}{z}
        -\delta_{iz}\frac{\gamma_{jz}-\delta_{jz}\gamma_{ww}}{z^2}
        +\frac{\partial_j\gamma_{iz}-\delta_{iz}\partial_j\gamma_{ww}}{z}
        -\delta_{jz}\frac{\gamma_{iz}-\delta_{iz}\gamma_{ww}}{z^2} \nn\\[10pt]
    &~~~~ \overset{*}{=} \left\{
    \begin{array}{ll}
      2\partial_z\partial_{(\hat{i}}\gamma_{\hat{j})z}~~~~&
          \text{if }i = \hat{i},~j = \hat{j} \\[10pt]
      0  & \text{if }(i,j)=(\hat{i},z) \text{ or }(z,\hat{j}) \\[10pt]
      \partial_z\partial_z(\gamma_{zz}-\gamma_{ww}) & \text{if }i=j=z
    \end{array}
    \right.\,.
  \end{align}

\item
  $$-\frac{1}{2} \frac{\partial_z \gamma_{ij}}{z}
        + \frac{\delta_{z(i} \gamma_{j)z} - \delta_{iz}
        \delta_{jz} \gamma_{ww}}{z^2}$$
  We use $\gamma_{zz}-\gamma_{ww}\overset{*}{=}\mathcal{O}(z^2)$
  and trade a division by $z$ for a $z$ derivative. The result is
  \begin{align}
    -\frac{1}{2} \frac{\partial_z \gamma_{ij}}{z}
        + \frac{\delta_{z(i} \gamma_{j)z} - \delta_{iz}
        \delta_{jz} \gamma_{ww}}{z^2}
    \overset{*}{=} &\left\{
    \begin{array}{ll}
      -\frac{1}{2} \partial_z \partial_z \gamma_{\hat{i}\hat{j}}~~~~~&
        \text{if }i=\hat{i},~j=\hat{j} \\[10pt]
      0 & \text{if }(i,j)=(\hat{i},z) \text{ or }(z,\hat{j}) \\[10pt]
      -\frac{1}{2} \partial_z \partial_z \gamma_{ww} & \text{if }i=j=z
    \end{array}
    \right.\,.\\\nonumber
  \end{align}

\item
  $$\frac{\gamma_{ww}\gamma^{zj}\partial_j\gamma_{ww}}{z}$$ \\
  Using Eqs.~(\ref{eq:gammamatrix})-(\ref{eq:invg}), we express
  the inverse metric components $\gamma^{zj}$ in terms of the
  downstairs metric and trade the division by $z$ for a $z$ derivative.
  We thus obtain
  \begin{align}
    \frac{\gamma_{ww}\gamma^{zj}\partial_j\gamma_{ww}}{z}
    \overset{*}{=} & \sum_{\hat{j}=1}^{d-1}\sum_{\hat{m}=1}^{d-1}(-1)^{\hat{m}
        +\hat{j}-1}\frac{\gamma_{ww}}{\det(M_{pq})}\partial_{\hat{j}}
        \gamma_{ww}\partial_z\gamma_{z\hat{m}}\det(M_{kl\{k \ne \hat{j},
        k \ne z, l \ne z, l \ne \hat{m}\}})
        \nonumber \\
    &+ \frac{\gamma_{ww}\det(M_{kl\{k\ne z, l\ne z\}})}{\det(M_{pq})}
        \partial_z\partial_z\gamma_{ww}\,.
  \end{align}
  which in the case $d=3$ reduces to
  \begin{align}
    \frac{\gamma_{ww}\gamma^{zj}\partial_j\gamma_{ww}}{z}
    \overset{*}{=} & \frac{(\gamma_{yx}\partial_z\gamma_{zy}-\gamma_{yy}
        \partial_z\gamma_{zx})\partial_x\gamma_{ww}+(\gamma_{yx}
        \partial_z\gamma_{zx}-\gamma_{xx}\partial_z\gamma_{yz})
        \partial_y\gamma_{ww}}{\gamma_{xx}\gamma_{yy}-\gamma_{xy}^2}
        \nonumber \\
    &+\partial_z\partial_z\gamma_{ww}\,.
  \end{align}

\item $$\frac{\gamma^{zz}\gamma_{ww}^2-\gamma_{ww}}{z^2}$$

The regularisation of this term proceeds in analogy to that of
term (9) in Appendix B of \cite{Cook:2016soy},
except we do not set $\det \gamma=1$. By rewriting
$1=\gamma^{zz}/\gamma^{zz}= \gamma^{zz} \det \gamma_{\I\J} / C_{zz}$,
trading divisions by $z$ for $z$ derivatives and using
$\gamma_{zz}\overset{*}{=}\gamma_{ww} + \mathcal{O}(z^2)$, we obtain
\begin{align}
  \frac{\gamma^{zz}\gamma_{ww}^2-\gamma_{ww}}{z^2}\overset{*}{=} &
  \frac{1}{2}\partial_z\partial_z(\gamma_{ww}-\gamma_{zz})
  \nn\\ &-\frac{\sum_{\hat{i}=1}^{d-1}\sum_{\hat{m}=1}^{d-1}(-1)^{\hat{i}
        +\hat{m}-1}\partial_z\gamma_{z\hat{i}}\partial_z\gamma_{z\hat{m}}
        \det(M_{kl\{ k\ne \hat{i}, k \ne z, l\ne z, l\ne \hat{m}\}})}
        {\det(M_{kl\{k\ne z, l \ne z\}})}\,. \\\nonumber
\end{align}
which in the case $d=3$ reduces to
\begin{align}
  \frac{\gamma^{zz}\gamma_{ww}^2-\gamma_{ww}}{z^2}\overset{*}{=} &
        \frac{1}{2}\partial_z\partial_z(\gamma_{ww}-\gamma_{zz})
        \nonumber \\
      & +\frac{-2\gamma_{xy}\partial_z\gamma_{xz}\partial_z\gamma_{yz}
        +\gamma_{xx}(\partial_z\gamma_{yz})^2
        +\gamma_{yy}(\partial_z\gamma_{xz})^2}
        {\gamma_{xx}\gamma_{yy}-\gamma_{xy}^2}\,. \\\nonumber
\end{align}
\item $$\frac{K_{iz} - \delta_{iz} K_{ww}}{z}$$

The division by $z$ is again traded for a derivative if
$i\ne z$ and for $i=z$, we use $K_{zz}=K_{ww}+\mathcal{O}(z^2)$, so that
\begin{align}
  \frac{K_{iz} - \delta_{iz} K_{ww}}{z}\overset{*}{=} &
  \left\{ \begin{array}{ll}
        \partial_z K_{\hat{i} z}~~~~~& \text{if }i = \hat{i} \\[10pt]
  0 & \text{if }i = z
\end{array}
\right.\,.
\end{align}

\end{list}

\section{Normalisation of the spatial normal frame vectors}\label{coordtrans}
In this section, we discuss how the set of spatial normal frame vectors,
Eq. (\ref{eq:mtilde}), can be recast in a form suitable for applying Gram-Schmidt
orthonormalisation. It turns out to be convenient to first rescale the
$\tilde{m}_{(\alpha)}$ such that they would acquire unit length in a flat
spacetime with spatial metric $\delta_{\I\J}$. Denoting these rescaled vectors
with a caret, we have
\begin{equation}
  \hat{m}_{(\alpha)} =
  \frac{1}{\sqrt{\left({\displaystyle \sum_{s=\alpha}^{D-1}} w_{s}^2 \right)\,
                 \left({\displaystyle \sum_{s=\alpha-1}^{D-1}} w_{s}^2 \right)}}
  \left(
  \begin{matrix}
    \left.
    \begin{matrix}
    0~~ \\
    \vdots~~ \\
    0~~ \\
    \end{matrix}
    \right\} (\alpha-2) \times
    \\[25pt]
    -\sum_{s=\alpha}^{D-1} (w^{s})^2 \\[15pt]
    \left.
    \begin{matrix}
    w^{\alpha-1}w^{\alpha} \\
    \vdots \\
    w^{\alpha-1} w^{D-2} \\
    w^{\alpha-1} w^{D-1}
    \end{matrix}
    ~~\right\} (D-\alpha)\times
  \end{matrix}
  \right)\,,~~\alpha=2,\,\ldots,\,D-1\,.
  \label{eq:ma_v1}
\end{equation}
Recall that we formally set $w^1\equiv x^1,\,\ldots,\,w^{d-1}\equiv x^{d-1}$,
$w^d\equiv z$.
As a convenient shorthand, we define
\begin{equation}
  \rho_{\I}^2 \equiv \sum^{D-1}_{s=\I} (w^{s})^2\,,
\end{equation}
so that, for instance, $\rho_1^2=r^2$,~~$\rho_4^2
= (w^4)^2+\ldots+(w^{D-1})^2$,~~$\rho_{D-1} = w^{D-1}$.
This definition allows us to write
\begin{equation}
  \hat{m}_{(\alpha)}^{\I} =
  \frac{1}{\rho_{\alpha}\,\rho_{\alpha-1}}
  \left(
  \begin{matrix}
    \left.
    \begin{matrix}
      0~~ \\
      \vdots~~ \\
      0~~ \\
    \end{matrix}
    \right\} (\alpha-2) \times
    \\[25pt]
    -\rho_{\alpha}^2 \\[10pt]
    \left.
    \begin{matrix}
      w^{\alpha-1}w^{\alpha} \\
      \vdots \\
      w^{\alpha-1} w^{D-2} \\
      w^{\alpha-1} w^{D-1}
    \end{matrix}
    ~~\right\} (D-\alpha)\times
  \end{matrix}
  \right)\,.
  \label{eq:ma_v2}
\end{equation}
We can now express the angles $\phi^{\alpha}$
in terms of the radial variables $\rho_{\I}$,
\begin{equation}
  \sin \phi^{\alpha} = \frac{\rho_{\alpha}}{\rho_{\alpha-1}}\,,~~~~~~~
  \cos \phi^{\alpha} = \frac{w^{\alpha-1}}{\rho_{\alpha-1}}\,.
\end{equation}
Using these relations in (\ref{eq:ma_v2}), we obtain
\begin{equation}
  \hat{m}_{(\alpha)}^{\I} =
  \left(
  \begin{matrix}
    \left.
    \begin{matrix}
      0~~ \\
      \vdots~~ \\
      0~~ \\
    \end{matrix}
    \right\} (\alpha-2) \times
    \\[25pt]
    -\sin \phi^{\alpha} \\[10pt]
    \left.
    \begin{matrix}
      \cos \phi^{\alpha} \,\cos \phi^{\alpha+1} \\
      \cos \phi^{\alpha} \,\sin \phi^{\alpha+1} \,\cos \phi^{\alpha+2} \\
      \vdots \\
      \cos \phi^{\alpha} \,\sin \phi^{\alpha+1} \,\ldots\,
      \sin \phi^{D-2} \,\cos \phi^{D-1} \\
      \cos \phi^{\alpha} \,\sin \phi^{\alpha+1} \,\ldots\,
      \sin \phi^{D-2} \,\sin \phi^{D-1} \\
    \end{matrix}
    ~~\right.
  \end{matrix}
  \right)
  =
  \left(
  \begin{matrix}
    \left.
    \begin{matrix}
      0~~ \\
      \vdots~~ \\
      0~~ \\
    \end{matrix}
    \right\} (\alpha-2) \times
    \\[25pt]
    -\sin \phi^{\alpha} \\[10pt]
    \begin{matrix}
      \vdots\\
      \cos \phi^{\alpha} \left(
      {\displaystyle \prod_{s=\alpha+1}^{\alpha+n-1}}
      \sin \phi^{s} \right)
      \cos \phi^{\alpha+n} \\
      \vdots
    \end{matrix}
  \end{matrix}
  \right)\,,
  \label{eq:ma_v3}
\end{equation}
where $n = 1,\,\ldots,\, D-\alpha$, and we formally set
$\cos \phi^{D-1} \equiv 1$ and
$\prod_{s=\alpha+1}^{\alpha} \sin \phi^{\alpha} \equiv1$.

Now, in our computational domain $\rho_{d+1}^2=0$, which,
from the definition of our coordinate system in Eq. (\ref{eq:coordinates}) gives
\begin{equation}
  r^2 \sin^2 \phi^2\,\ldots\,\sin^2 \phi^{d+1}=0
\end{equation}
Since $\phi^2,\,\ldots,\,\phi^{d}$ are arbitrary in our computational domain,
we must have either $\phi^{d+1} =0 ~ \mathrm{or} ~ \pi$.
Without loss of generality, we choose $\phi^{d+1}=0$, which fixes the
$d-1$ vectors
\begin{align}
  \hat{m}_{(2)} =& (-\sin \phi^2,~\cos \phi^2 \,\cos\phi^3,
        ~\ldots,~
        \cos \phi^2
        {\displaystyle \prod_{s=3}^d \sin (\phi^s)},\,
        \underbrace{0,~\ldots,~0}_{(D-d-1)\times})\,.
  \label{eq:hatm2} \\
  \vdots & \nonumber \\
  \hat{m}_{(\hat{\alpha})}
        =& (\underbrace{0,\,\ldots,\,0}_{(\hat{\alpha}-2)\times},
        \,-\sin \phi^{\hat{\alpha}},~\cos \phi^{\hat{\alpha}}
        \cos \phi^{\hat{\alpha}+1},\,
        \ldots,\,\cos \phi^{\hat{\alpha}}
        {\displaystyle\prod_{s=\hat{\alpha}+1}^d} (\sin \phi^s),\,
        \underbrace{0,\,\ldots,\,0}_{(D-d-1)\times})
        \\
  \vdots & \nonumber \\
  \hat{m}_{(d)} =& (\underbrace{0,\,\ldots,\,0}_{(d-2)\times},\,
        ~-\sin \phi^d,~\cos \phi^d,~
        \underbrace{0,~\ldots,~0}_{(D-d-1)\times})\,,
  \label{eq:hatm3}
\end{align}
which, up to rescaling by $\rho_{\hat{\alpha}} \rho_{\hat{\alpha}-1}$,
are equal to the vectors in Eqs.~(\ref{eq:tildem2gen})-(\ref{eq:tildemdgen}).
For the remaining vectors, we can use the rotational freedom in the angles
$\phi^{d+2},\,\ldots,\,\phi^{D-1}$. Any choice for these values will satisfy
$w^{d+1}=\ldots = w^{D-1}=0$
as required on our computational domain and we merely
need to ensure that we choose these angles such that the resulting set of
vectors is orthogonal. This is most conveniently achieved by setting
\begin{equation}
  \phi^{d+2} = \ldots = \phi^{D-1} = 0\,,
\end{equation}
which, inserted into Eq.~(\ref{eq:ma_v3}), implies
\begin{equation}
  \hat{m}_{(a)}^{\I}=\delta^{\I}_{a},\,~~~~~a=d+1,\ldots,D-1.
\end{equation}
Combined with Eqs.~(\ref{eq:hatm2})-(\ref{eq:hatm3}) and restoring the
tilde in place of the caret on the $\tilde{m}_{(a)}$, we have recovered
Eqs.~(\ref{eq:tildemdp1gen})-(\ref{eq:tildemDm1gen})
in Section \ref{mvecs} for the angular vectors.
For the case $d=3$ we have just two non-trivial vectors:
\begin{align}
  \hat{m}_{(2)} =& (-\sin \phi^2,~\cos \phi^2 \,\cos\phi^3,~\cos
        \phi^2\,\sin \phi^3,~
        \underbrace{0,~\ldots,~0}_{(D-4)\times})\,,\\
  \hat{m}_{(3)} =& (0,~-\sin \phi^3,~\cos \phi^3,~
        \underbrace{0,~\ldots,~0}_{(D-4)\times})\,,
\end{align}
recovering Eqs.~(\ref{eq:tildem2})-(\ref{eq:tildemDm1})

\section*{References}

\bibliographystyle{unsrt}

\end{document}